\documentclass[letterpaper,twocolumn,english,prl]{revtex4}

\usepackage{amsmath,amssymb,amsthm}
\usepackage{hyperref}
\usepackage{amsthm}
\usepackage{graphicx}
\usepackage{subfigure}
\usepackage{color,bbm}
\usepackage{amsfonts}
\usepackage{mathrsfs}
\usepackage{pdfpages}

\newcommand{\bA     }{\mbox{\boldmath$A$}}
\newcommand{\bC     }{\mbox{\boldmath$C$}}

\begin{document}
\date{\today}
\title{\bf  Localization and universality of eigenvectors in directed random graphs}

\author{Fernando Lucas Metz}
 \affiliation{Physics Institute, Federal University of Rio Grande do Sul, 91501-970 Porto Alegre, Brazil}
  \affiliation{London Mathematical Laboratory, 18 Margravine Gardens, London W6 8RH, United Kingdom}
\author{Izaak Neri} 
\affiliation{Department of Mathematics, King’s College London, Strand, London, WC2R 2LS, UK}
\begin{abstract}      
  Although the spectral properties of  random graphs have been a long-standing focus of network theory, the properties of  right eigenvectors of directed graphs have so far eluded an exact analytic treatment.
  We present a general theory for the  statistics of the  right eigenvector components in  directed random graphs with a prescribed degree distribution and with randomly weighted links.   We obtain exact analytic expressions for the  inverse participation ratio and show that   right eigenvectors of directed random graphs with a small average degree are localized.  Remarkably, if the fourth moment of the degree distribution is finite, then the critical mean degree of the localization transition is independent of the degree fluctuations, which is different from localization in undirected graphs that is governed by degree fluctuations.  We also show that in the high connectivity limit the distribution of the right eigenvector components is solely determined by the degree fluctuations.  For delocalized eigenvectors, we recover the universal results from standard random matrix theory that are independent of the degree distribution, while for localized eigenvectors the  eigenvector distribution depends on the degree distribution.  
 \end{abstract}

\maketitle
  
\paragraph{Introduction.}   
Complex systems, such as, neural networks \cite{brunel2000dynamics, bullmore2009complex, sporns2010networks}, ecosystems  \cite{bascompte2009disentangling}, gene regulatory networks \cite{milo2002network, shen2002network, lee2002transcriptional},  social networks \cite{kwak2010twitter, aiello2012friendship}, and the World Wide Web \cite{broder2000graph, pastor2007evolution} are described by
large, directed graphs. Therefore, there is much interest in understanding how  the topology of directed graphs impacts the dynamics of processes and algorithms on them.

Much insight in the dynamical processes on graphs is gained from the spectral properties of the adjacency matrix that represents the network.   This is because the linearized dynamics of a complex system in the vicinity of a fixed point is determined by the spectral properties of the adjacency matrix \cite{hartman1960lemma, grobman1959homeomorphism}.     As a consequence,  spectral analysis of the adjacency
matrix has proven to be important in the study of neural networks \cite{sompolinsky1988chaos, del2013synchronization, kadmon2015transition, aljadeff2015transition, marti2018correlations}, ecosystems \cite{May, Allesina2015, gibbs2018effect}, gene regulatory networks \cite{Chen2019, Guo2020}, and disease spreading \cite{Van2012,Gol2012,Pastor2018,Li2013, silva2020dissecting}. In these systems, the eigenvectors of the adjacency matrix determine the dynamical modes evoked by external perturbations.  In addition, right eigenvectors of adjacency matrices of directed graphs are used in  algorithms for node centrality \cite{Bona1972,Rest2006a, Travis2014}, community detection  \cite{krzakala2013spectral, bordenave2015non, kawamoto2018algorithmic}, and matrix completion  \cite{bordenave2020detection}.    
  
In disordered systems, eigenvectors  localize when the strength of the disorder is large enough \cite{abou1973selfconsistent, aizenman2011extended}. Localized eigenvectors occupy a few vertices, whereas delocalized eigenvectors are extended over the whole system. The transition from a delocalized to a localized state leads to a qualitative change in the dynamics of processes and algorithms. For example, the localization transition implies a metal-insulator phase transition in solid state physics \cite{abou1973selfconsistent, aizenman2011extended}, a transition from an algorithmic  successful  to a failure phase in  spectral algorithms    \cite{PhysRevE.80.026107, Travis2014, bordenave2020detection}, and a transition from a regime where the linear dynamics of a large complex system  is governed by a finite number of vertices
to a regime where the dynamics is governed by a finite fraction of all vertices. In the context of disease spreading, eigenvector localization
implies that the fraction of infected vertices is very small right above the epidemic threshold \cite{Gol2012}.

For undirected random graphs, the localization of eigenvectors of the adjacency matrix has been well studied~\cite{abou1973selfconsistent, Fyodorov1991, mirlin1991localization, evangelou1992numerical, bauer2001random, Kuhn2008, metz2010localization, aizenman2011extended, Gol2012, Kabashima2012, slanina2012, Pastor2016, tikhonov2016anderson, Pastor2018, Susca2019}.  The eigenvector of the largest eigenvalue is localized if the maximal degree of the graph is larger than a certain value.  Hence, degree fluctuations are crucial for the localization of eigenvectors in undirected graphs.

For directed random graphs, the statistical properties and the localization of eigenvectors have been studied for one-dimensional chains, such as, the Hatano-Nelson model \cite{hatano1996localization, hatano1997vortex,feinberg1999non} and its extensions to biological systems \cite{Amir2016, Zhang2019}, and  a diluted  Ginibre ensemble \cite{Peron}. However, the localization of eigenvectors in directed random graphs  that model complex systems, such as, the World Wide Web or neural networks, have not been studied so far.   
 
In this Letter, we make a significant step forward by developing an exact theory for the statistical properties of the right (or left) eigenvectors of directed random graphs with a prescribed degree distribution and random couplings. We derive exact analytic expressions for the inverse participation ratio and for the critical point of the localization-delocalization transition.
Surprisingly, when the moments of the degree distribution are finite, the critical point of the  localization-delocalization  transition is independent of the degree distribution. Moreover, the right eigenvectors  are localized if the degree distribution has a diverging fourth moment.
We also show that in the high connectivity limit the statistics of the components of right eigenvectors are only determined by degree fluctuations. In this limit, we obtain distinct universality classes that depend on an exponent that quantifies the degree fluctuations.

\paragraph{Model set-up.}   
We consider random matrices $\mathbf{A}$ of dimension $n\times n$ with elements 
\begin{equation}
A_{ij} = J_{ij}C_{ij}, \quad    i,j\in \left\{1,2,\ldots,n\right\}, \label{eq:Def}
\end{equation}
where $C_{ij}\in \left\{0,1\right\}$ are the entries of the adjacency matrix $\bC$ of  a simple and directed random graph with a
prescribed degree distribution 
\begin{eqnarray}
  p_{K^{\rm in}, K^{\rm out}}(k,\ell) = p_{K^{\rm in}}(k) p_{K^{\rm out}}(\ell)
  \label{ghaq}
\end{eqnarray}
of indegrees $K^{\rm in}$ and outdegrees $K^{\rm out}$. We set $C_{ij} = 1$ when there exists a directed link
pointing from $i$ to $j$, such that the outdegree  (indegree) of the $i$-th node is  $K^{\rm out}_i = \sum^n_{j=1}C_{ij}$ ($K^{\rm in}_i = \sum^n_{j=1}C_{ji}$).
The $J_{ij}$ are  real-valued, independent and identically distributed random variables drawn from a distribution $p_J(x)$.

Random graph models with undirected edges and a prescribed degree distribution are surveyed in \cite{Fosdick}. Here we consider their extension to the directed case. Directed
random graphs with a prescribed degree distribution  \cite{Molloy95,Molloy98, bollobas2001random, Newman2001,Newman10, dorogovtsev2013evolution}  model
the  World Wide Web \cite{broder2000graph, pastor2007evolution} and neural networks \cite{brunel2000dynamics, amari2003handbook, sporns2010networks}. In this model, the
indegrees and outdegrees are drawn from  Eq.~(\ref{ghaq}) subject to the  constraint $\sum^n_{j=1}K^{\rm in}_j = \sum^n_{j=1}K^{\rm out}_j $, and subsequently nodes
are randomly connected according to the given degree sequences.
Hence, given a sequence of degrees, random graphs are drawn uniformly from the set of simple and directed graphs.
This  model provides the ideal setting to explore
the influence of network topology on the spectral properties of $\bA$.     
  
In what follows,  brackets $\langle \cdot \rangle$ denote the average  with respect to the distribution of $\bA$. In particular, we use 
    \begin{equation}
    c = \langle K^{\rm out}\rangle 
  \end{equation}
  for the mean outdegree, and we denote the variance of a  random variable $X$ by ${\rm var}(X) = \langle X^2\rangle - \langle X\rangle^2$.

\begin{figure}
  \begin{center}
    \includegraphics[scale=0.6]{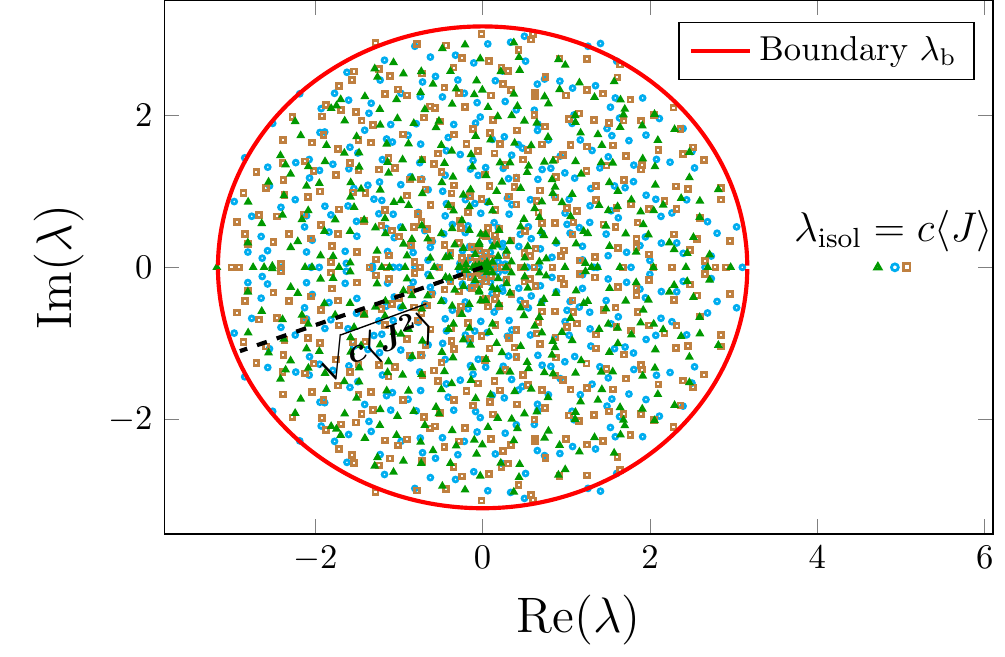} 
    \caption{Eigenvalues of three realizations (circles, triangles, and squares) of the adjacency matrix $\bA$ of directed random graphs with $n=500$ (see Eq.~(\ref{eq:Def})). The indegrees
      and outdegrees follow a Poisson distribution with average $c=5$. The weights $J_{ij}$ are
      drawn from a Gaussian distribution $p_J$ with mean and variance equal to one.
}
\label{spec}
\end{center}
\end{figure}

\paragraph{Spectra of infinitely large matrices $\mathbf{A}$.}
The spectrum of $\mathbf{A}$ has been studied  in  Refs.~\cite{Rogers2009, Neri2016, Metz2019, Neri2019}.   
For $n \rightarrow \infty$ and $c>1$, directed random graphs have a giant strongly connected component~\cite{dorogovtsev2001giant} and the spectral distribution
$\rho_{\mathbf{A}}(\lambda) = n^{-1}\sum^n_{j=1}\delta( \lambda-\lambda_j(\mathbf{A}))$ of the eigenvalues $\{ \lambda_j(\mathbf{A}) \}_{j=1}^n$ is
supported on a disk of radius $|\lambda_{\rm b}| = \sqrt{c \langle J^2 \rangle}$ centered at the origin of the complex plane. In addition, if
\begin{equation}
c >c_{\rm gap}= \frac{ \langle J^2 \rangle}{\langle J \rangle^2}, \label{eq:gap}
\end{equation}
then there exists an eigenvalue outlier located at $\lambda_{\rm isol} = c \langle J \rangle$ that is separated
from the boundary $\lambda_{\rm b}$ by a finite gap.   
Figure~\ref{spec} shows the eigenvalues for an example of a directed random graph, where one clearly identifies
the outlier $\lambda_{\rm isol}$ and the boundary $\lambda_{\rm b}$ of $\rho_{\mathbf{A}}(\lambda)$ for $n \rightarrow \infty$.

\paragraph{Distribution of the right eigenvector components.}
A  right  eigenvector $\vec{R}(\lambda)$ associated to an eigenvalue $\lambda$ of $\bA$ satisfies
\begin{equation}
  \bA\vec{R}(\lambda) = \lambda \vec{R}(\lambda),
\end{equation}
and the distribution of the entries of $\vec{R}(\lambda)$ reads
\begin{equation}
p_R(r|\lambda) = \lim_{n \rightarrow \infty} \frac{1}{n}\sum^n_{i=1}\delta\left(r -R_i(\lambda) \right).
\end{equation}
If $\lambda$ is an outlier ($\lambda = \lambda_{\rm isol}$) or $\lambda$ is located at the boundary of the spectrum ($\lambda = \lambda_{\rm b}$), then $p_R(r|\lambda)$
fulfills \cite{Neri2016, Metz2019, Neri2019} 
\begin{align}
  p_R(r|\lambda) &= \sum_{k=0}^{\infty} p_{K^{\rm out}}(k) \int \left( \prod_{j=1}^{k} {\rm d} x_j {\rm d}^2 r_j p_J(x_j) p_R(r_j|\lambda)    \right) \nonumber \\
 &\times \delta\left(r - \frac{1}{\lambda} \sum_{j=1}^k x_j r_j   \right),
  \label{gh1} 
\end{align}  
where ${\rm d}^2 r \equiv {\rm d} \, {\rm Re} \, r \,\, {\rm d} \, {\rm Im} \, r $. Equation (\ref{gh1}) is
exact for infinitely large and directed random graphs with a prescribed degree distribution, because they are locally tree-like. In fact, the solutions
of Eq.~(\ref{gh1}) are well corroborated by  
direct diagonalizations of large adjacency matrices \cite{Neri2016,Metz2019,Neri2019}. The analytic results presented below
follow  from Eq.~(\ref{gh1}).

\paragraph{Inverse participation ratio.}
The localization of $\vec{R}(\lambda)$ can be  characterized in terms of the inverse participation ratio (IPR)~\cite{Mirlin1994, efetov1999supersymmetry,metz2010localization}
\begin{equation}
  \mathcal{I}(\lambda) \equiv \lim_{n\rightarrow\infty}  \frac{n \sum_{i=1}^n |R_{i}(\lambda)|^4}{\left( \sum_{i=1}^n |R_{i}(\lambda)|^2\right)^2}
  = \frac{\langle |R(\lambda)|^4 \rangle}{\langle |R(\lambda)|^2 \rangle^2},  
\end{equation}   
where we have used that $\mathcal{I}$ is self-averaging \cite{SM}. The IPR is finite if $\vec{R}(\lambda)$ is delocalized, whereas $\mathcal{I}(\lambda)$
diverges if $\vec{R}(\lambda)$ is localized on a finite number of nodes.
 
From  Eq.~(\ref{gh1}), we derive in the Supplemental Material~\cite{SM} exact expressions for the IPR when $\lambda=\lambda_{\rm isol}$ or $\lambda = \lambda_{\rm b}$.  We find that  
\begin{equation}
  \mathcal{I}(\lambda_{\rm b})
  = \frac{\left( \gamma + 1 \right)   \left[ \langle (K^{\rm out})^2 \rangle -c \right] }
       {c \left(c  - \langle  J^4  \rangle/\langle J^2  \rangle^2  \right)},
       \label{bound}
\end{equation}
where  $\gamma = 2$  when $\lambda_{\rm b} \in \mathbb{R}$  and $\gamma=1$ when
$\lambda_{\rm b} \notin \mathbb{R}$. Analogously, the IPR at $\lambda=\lambda_{\rm isol}$ reads
\begin{align}
 \mathcal{I}(\lambda_{\rm isol}) &= \frac{3 \beta_1 \langle J^2 \rangle^2  }
       {\left( c^4 \langle J \rangle^4 - c \langle J^4 \rangle \right)} +  \frac{\beta_3 \left( c^2 \langle J \rangle^2 - c \langle J^2 \rangle \right)^2 }
       {\beta_1^2 \left( c^4 \langle J \rangle^4 - c \langle J^4 \rangle \right)  } 
       \nonumber \\
       &+ \frac{ 12 \beta_1 \langle J^3 \rangle  \langle J^2 \rangle \left( c^2 \langle J\rangle^2 - c \langle J^2 \rangle \right)  }
                 {   \left( c^4 \langle J \rangle^4 - c \langle J^4 \rangle \right)  \left( c^3 \langle J \rangle^3 - c \langle J^3 \rangle \right)   } \nonumber \\
                 &+  \frac{4 \beta_2 \langle J^3 \rangle \left( c^2 \langle J \rangle^2 - c \langle J^2 \rangle    \right)^2  }
                 {\beta_1    \left( c^4 \langle J \rangle^4 - c \langle J^4 \rangle \right)  \left( c^3 \langle J \rangle^3 - c \langle J^3 \rangle \right)   } \nonumber \\
                 &+ \frac{6 \beta_2 \langle J^2 \rangle   \left( c^2 \langle J \rangle^2 - c \langle J^2 \rangle \right)    }
                 {\beta_1  \left( c^4 \langle J \rangle^4 - c \langle J^4 \rangle \right)   }\,,
                 \label{out}
\end{align}   
 where 
\begin{equation}
\beta_{\ell} \equiv  \sum^{\infty}_{k=\ell+1}p_{K^{\rm out}}(k) \frac{k!}{\left(k - \ell -1  \right)!}  ,\quad \ell = 1,2,3.
\end{equation}  

\begin{figure}
  \begin{center}
     \includegraphics[scale=0.72]{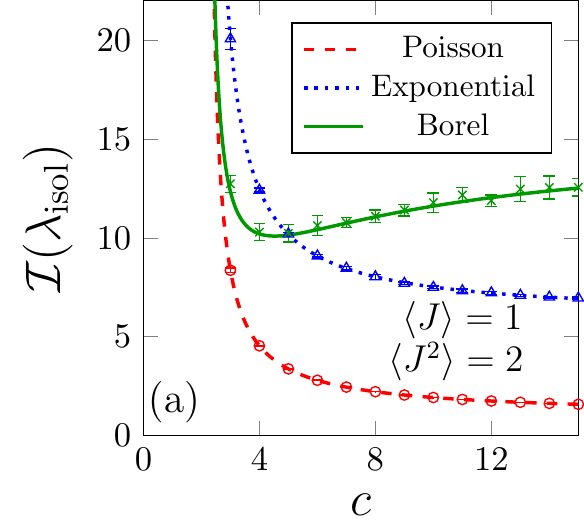}
     \includegraphics[scale=0.72]{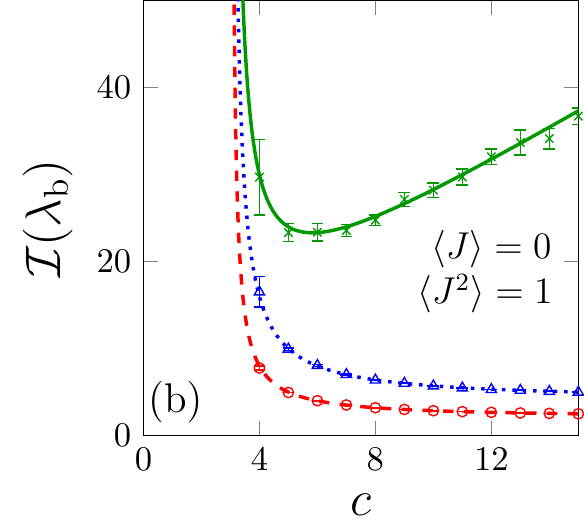}
     \caption{The  IPR $\mathcal{I}(\lambda)$ of right eigenvectors
       associated to $\lambda_{\rm isol}$ [Panel (a)] and $\lambda_{\rm b} \notin \mathbb{R}$ [Panel (b)]. Equations (\ref{bound}) and (\ref{out})
    (different line styles) are shown as a function of the average degree $c$ for different outdegree distributions: Poisson, exponential, and Borel (see Supplemental Material \cite{SM}).
    The weights $J_{ij}$ are drawn from a Gaussian distribution $p_J$ with first and second
    moments indicated on each panel. The symbols are obtained from the numerical solutions
    of Eq. (\ref{gh1})  using the population dynamics algorithm \cite{Kuhn2008,Metz2019}, while direct diagonalization results
    for $\mathcal{I}(\lambda)$ are presented in the Supplemental Material \cite{SM}.
The error bars are the standard deviation of the IPR for 10 independent runs of population dynamics.
    The results for the Borel distribution are rescaled
    as $\mathcal{I}(\lambda_{\rm isol}) \rightarrow \mathcal{I}(\lambda_{\rm isol})/c$ in panel (a).  
}
\label{IPR}
\end{center}
\end{figure}

Figure \ref{IPR} illustrates Eqs.~(\ref{bound}) and (\ref{out}) as a function of $c$ for a Gaussian distribution $p_J$ and three different outdegree distributions: Poisson, exponential, and
Borel distribution (see Supplemental Material \cite{SM}). All moments of these degree distributions are finite and each $p_{K^{\rm out}}$ is parametrized only by $c$.
Figure \ref{IPR} shows  that the IPR is finite if $c$ is large enough and it diverges for small $c$, which  demonstrates the existence of a delocalization-localization phase transition in directed random graphs.    
 
\paragraph{The localization phase transition.} 
There are two mechanisms for localization, one governed by fluctuations of $J_{ij}$, and a second one governed by degree fluctuations.   

The first mechanism  is illustrated in Fig.~\ref{IPR} and it holds for arbitrary $p_{K^{\rm out}}$ with a finite fourth moment.
In this case, the right eigenvectors associated to $\lambda = \lambda_{\rm b}$   and $\lambda=\lambda_{\rm isol}$  are localized when  $c$ is smaller than 
 \begin{equation}
   c_{\rm b} = \frac{\langle J^4 \rangle}{\langle J^2 \rangle^2} \quad {\rm and} \quad  c^3_{\rm isol} = \frac{\langle J^4 \rangle}{\langle J \rangle^4},
   \label{critic}
  \end{equation}  
 respectively. Thus, the critical points for the localization transitions only depend on the lower
 moments of $p_J$ and they are independent of $p_{K^{\rm out}}$. When $p_J(x) = \delta(x-1)$,  we obtain $c_{\rm b}=c_{\rm isol}=1$ and
 the delocalization-localization transition is governed
 by the  percolation transition for the strongly connected component \cite{dorogovtsev2001giant}.
 According to Eq.~(\ref{out}), a localization transition at $c^\ast_{{\rm isol}} = \sqrt{ \langle J^3\rangle/\langle J \rangle^3} $ is in
 principle possible, but we could not find an example of $p_J$ for which $c^\ast_{\rm isol}>c_{\rm isol}$ and $c^\ast_{\rm isol} >c_{\rm gap}$. 

\begin{figure}
  \begin{center}
    \includegraphics[scale=0.65]{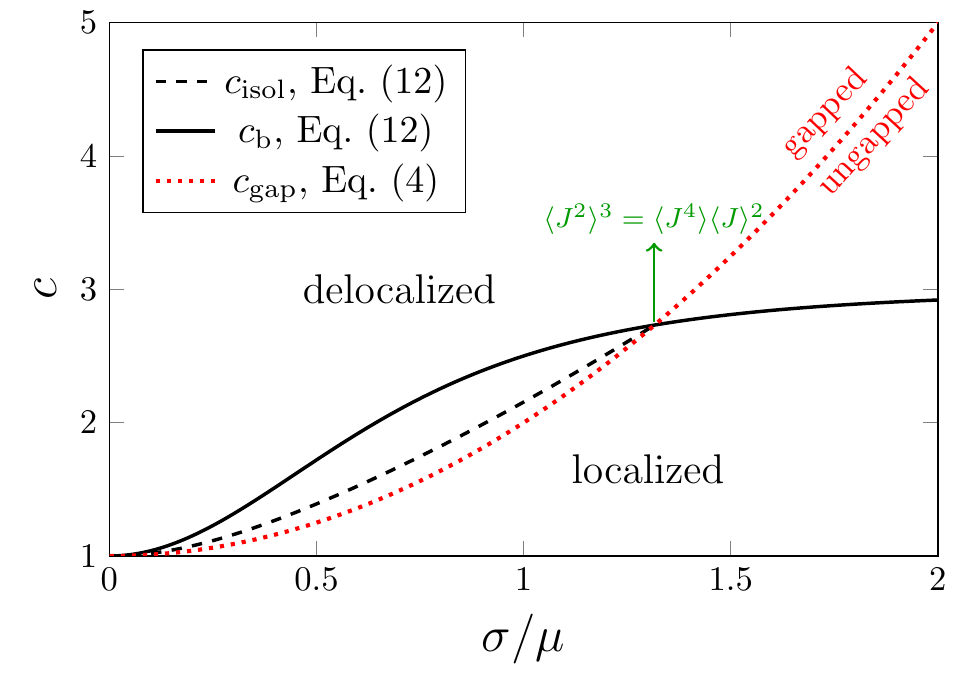} 
    \caption{Phase diagram for the localization  of  right eigenvectors  associated to $\lambda_{\rm isol}$ and $\lambda_{\rm b}$. The distribution $p_J$ is Gaussian  with mean $\mu$ and
      standard deviation $\sigma$.
}
\label{specLoc}
\end{center}
\end{figure}

 Figure \ref{specLoc} shows the phase diagram when 
 $p_J$ is a Gaussian distribution with mean $\mu$ and variance $\sigma^2$.  In this case,  $c_{\rm gap}$, $c_{\rm b}$ and $c_{\rm isol}$ only depend on $\sigma/\mu$.
 A few generic properties of eigenvector localization in directed random graphs, which also hold for non-Gaussian $p_J$, are illustrated in Fig.~\ref{specLoc}. First, $\vec{R}(\lambda_{\rm isol})$
 is delocalized  when $\langle J^2\rangle^3 > \langle J^4\rangle\langle J \rangle^2$ because $c_{\rm gap}>c_{\rm isol}$. Second, the transition lines
 fulfill $c_{\rm gap} < c_{\rm isol} < c_{\rm b}$ for $\langle J^2\rangle^3 < \langle J^4\rangle\langle J \rangle^2$.    Lastly, the critical transitions $c_{\rm gap}$, $c_{\rm isol}$ and $c_{\rm gap}$ intersect
 in a common point because $c_{\rm isol}^3 = c_{\rm b} c_{\rm gap}^2$.  
 
 The second mechanism for localization is due to large degree fluctuations. From Eqs.~(\ref{bound}) and (\ref{out}), it follows that $\mathcal{I}(\lambda_{\rm b}) \rightarrow \infty$ if $\langle (K^{\rm out})^2\rangle \rightarrow \infty$ and $\mathcal{I}(\lambda_{\rm isol}) \rightarrow \infty$ if $\langle (K^{\rm out})^4\rangle \rightarrow \infty$, independently of $p_J$. Hence, localization of $\vec{R}(\lambda_{\rm b})$ and $\vec{R}(\lambda_{\rm isol})$ also occurs in graphs with power-law degree distributions. In the sequel, we show that degree-based localization persists in the high connectivity limit.
 
\paragraph{Localization and universality in the high connectivity limit.}  
In Fig.~\ref{IPR}, $\mathcal{I}(\lambda)$ flows to different asymptotic values
for $c \gg 1$. To explore the localization and universality of eigenvectors in the high connectivity limit  $c \rightarrow \infty$, we
analyze the  moments of the distribution $p_R$. Since $\langle R(\lambda_{\rm isol}) \rangle$ is finite, we characterize the
limit $c \rightarrow \infty$ of $p_R(r|\lambda_{\rm isol})$ through the relative variance 
\begin{equation}
  \mathcal{R}_c = \frac{{\rm var}[R(\lambda_{\rm isol})]}{\left\langle   R(\lambda_{\rm isol})  \right\rangle^2 }.  
  \label{kaqpa}
\end{equation}  
On the other hand, since  $\langle R(\lambda_{\rm b} )\rangle=0$, we characterize 
 the limit  $c \rightarrow \infty$ of $p_R(r|\lambda_{\rm b})$ through the kurtosis
\begin{equation}
  \mathcal{K}_c = \frac{\left\langle  \left( {\rm Re}\: R(\lambda_{\rm b})     \right)^4     \right\rangle }
          {\left\langle \left( {\rm Re}\: R(\lambda_{\rm b})     \right)^2   \right\rangle^2 } =
\frac{\left( 4 - \gamma \right)}{2} \mathcal{I}(\lambda_{\rm b})   ,       \label{eq:Kc}
\end{equation}   
where we used the fact that odd moments of $p_R(r|\lambda_{\rm b})$ are zero \cite{SM}.
Setting $c \rightarrow \infty$ in Eqs.~(\ref{kaqpa}) and (\ref{eq:Kc}), we obtain \cite{SM}
\begin{align}
\mathcal{R}_\infty &= \lim_{c \rightarrow \infty} \frac{{\rm var}[K^{\rm out}]}{c^2}, \label{uni1}\\ 
\mathcal{K}_\infty &= 3 \left( 1 + \lim_{c \rightarrow \infty} \frac{{\rm var}[K^{\rm out}]}{c^2} \right), \label{uni2} 
\end{align}  
which indicates that the limit $c \rightarrow \infty$ of $p_R$ is determined by the degree distribution. We see that, in general, $p_R(r|\lambda_{\rm b})$ and $p_R(r|\lambda_{\rm isol})$ are not
Gaussian in the high connectivity limit.

With the purpose of classifying the universal behavior of $p_R$ for $c \rightarrow \infty$, let us consider degree distributions
that satisfy
\begin{equation}
{\rm var}[K^{\rm out}] = B c^{\alpha} \quad (c \gg 1), \label{hapqo}
\end{equation}
where  $\alpha$ and 
$B$ depend on the specific  choice of $p_{K^{\rm out}}(k)$. Equation (\ref{hapqo}) holds for  different
  examples of degree distributions, including those  in Fig.~\ref{IPR}.   Plugging this {\it ansatz} for ${\rm var}[K^{\rm out}]$ in Eqs.~(\ref{uni1}) and (\ref{uni2}), we obtain
three  universality classes for $\lim_{c \rightarrow \infty} p_R(r|\lambda)$, which are determined by the exponent $\alpha$ that controls the degree fluctuations.
The results for the universality classes are summarized in table \ref{univ}.  In terms of $\mathcal{R}_\infty$ and $\mathcal{K}_\infty$,
we find that for $\alpha \leq 2$ the eigenvectors $\vec{R}(\lambda_{\rm b})$
and $\vec{R}(\lambda_{\rm isol})$ are delocalized in the limit $c \rightarrow \infty$, whereas for $\alpha> 2$ these eigenvectors are localized due to large degree fluctuations.    

\begin{table}[ht]
\centering 
\begin{tabular}{c c c c} 
\hline\hline 
 & $\alpha < 2$ & $\alpha=2$ & $\alpha > 2$ \\ [0.5ex] 
\hline 
$\mathcal{R}_{\infty}$ & $0$ &  $B$ &  $\infty$ \\
$\mathcal{K}_{\infty}$ & $3$  & $3 (1+B)$   & $\infty$ \\ 
Example & Poisson & Exponential & Borel \\
[1ex] 
\hline 
\end{tabular}
\caption{The relative variance  $\mathcal{R}_{c}$ of  $\vec{R}(\lambda_{\rm isol})$ and the kurtosis $\mathcal{K}_{c}$ of  $\vec{R}(\lambda_{\rm b})$ in the high connectivity limit $c\rightarrow \infty$ (see Eqs.~(\ref{uni1}) and (\ref{uni2})), together with an example of the outdegree distribution $p_{K^{\rm out}}$ in each regime of $\alpha$ (see Eq.~(\ref{hapqo})).}
\label{univ} 
\end{table}

\paragraph{The eigenvector distributions in the high connectivity limit.}  
The results in Table~\ref{univ} indicate that  $p_R(r|\lambda)$ is universal for $c \rightarrow \infty$. Below we present explicit
expressions for $p_R(r|\lambda)$ when $c \rightarrow \infty$. Henceforth we
set $\langle |R|^2 \rangle = 1$ without loosing generality.

The  characteristic function of $p_R(r|\lambda)$ is given by \cite{SM}
\begin{equation}
  g_R(u,v|\lambda) = \sum_{k=0}^{\infty} p_{K^{\rm out}}(k) e^{k \ln{F(u,v|\lambda) }},
  \label{lla}
\end{equation}  
where 
\begin{equation}
  F(u,v|\lambda) = \int {\rm d}x \: p_J(x) \int {\rm d}^2r \, p_R(r|\lambda)   e^{- \frac{x z r}{2 \lambda}   +  \frac{x z^* r^*}{2 \lambda^*}   }  ,
  \label{hj}
\end{equation}
and $z = u + i v$. The symbol $(\dots)^*$ denotes complex-conjugation. 
If $\lambda \in \mathbb{R}$, the eigenvector components are real and $F(u,v|\lambda)$ is independent of $v$.  

Setting $\lambda = \lambda_{\rm isol}$ or $\lambda = \lambda_{\rm b}$ in Eq.~(\ref{hj}), we can expand $F(u,v|\lambda)$ for $c \gg 1$ up to order $O(1/c)$ if  $\alpha \leq 2$
(see table \ref{univ}).
This approach does not work for $\alpha>2$, because  the  moments of $p_R$ can diverge in this regime. Thus, performing this expansion for $\alpha \leq 2$ and substituting the resulting expression
for $F(u,v|\lambda)$ in Eq. (\ref{lla}), we obtain \cite{SM}
\begin{align}
&g_R(u,v|\lambda_{\rm b}) = \sum_{k=0}^{\infty} p_{K^{\rm out}}(k) \exp{\left[   - \frac{\gamma k}{4 c} \left( u^2 + \left( 2 - \gamma  \right) v^2 \right) \right]}, \label{hja1} \\
&g_R(u,v|\lambda_{\rm isol}) = \sum_{k=0}^{\infty} p_{K^{\rm out}}(k) \exp{\left( - \frac{{\rm i} u k}{c \sqrt{B c^{\alpha-2} + 1}}    \right)}. \label{hja2}
\end{align}  
Remarkably,  the characteristic functions for $c \rightarrow \infty$ are
fully specified by $p_{K^{\rm out}}$ and they are independent of~$p_J$.

For degree distributions where $\lim_{c \rightarrow \infty} {\rm var}[K^{\rm out}]/c^2=0$ ($\alpha < 2$), it is reasonable to set $p_{K^{\rm out}}(k) = \delta_{k,c}$ in
Eqs.~(\ref{hja1}) and (\ref{hja2}), leading to \cite{SM}
\begin{align}
&p_R(r|\lambda_b) = \frac{1}{\pi} e^{-|r|^2} \quad (\lambda_b\notin \mathbb{R}), \label{kop1} \\
&p_R(r|\lambda_{\rm isol}) = \delta\left[ {\rm Im} (r)  \right] \delta\left[ {\rm Re} (r) - 1  \right] . \label{eq:alpha2}
\end{align} 
Equation (\ref{kop1}) yields the  well-known  Porter-Thomas distribution for the eigenvector components of Gaussian random matrices \cite{Porter1956,Mirlin2000}. Thus, standard results
from random matrix theory are recovered when $\alpha<2$.
   
\begin{figure}
  \begin{center}
    \includegraphics[scale=0.9]{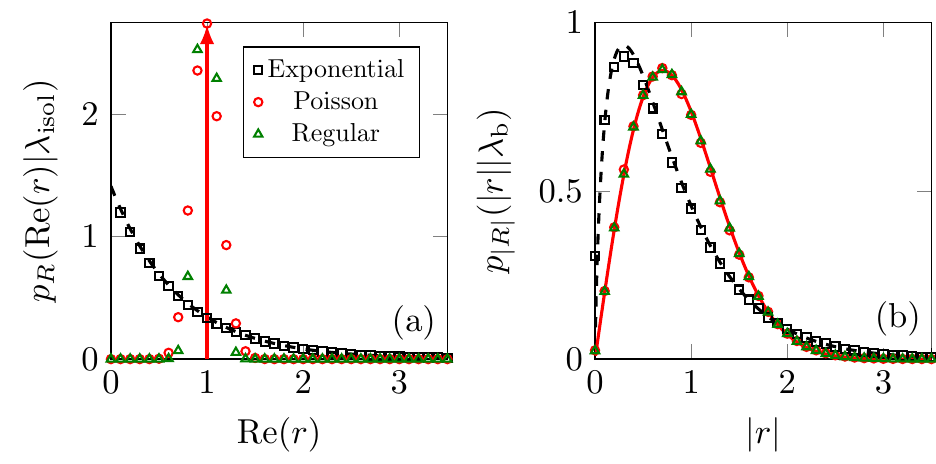}
    \caption{The high connectivity limit $c \rightarrow \infty$ of the distribution $p_R({\rm Re} (r)|\lambda_{\rm isol})$ of the real part of the
      eigenvector components at $\lambda_{\rm isol}$ [Panel (a)], and of the distribution $p_{|R|}(|r||\lambda_{\rm b})$ of the norm of the eigenvector
      components at $\lambda_{\rm b} \notin \mathbb{R}$ [Panel (b)]. The solid red lines and the dashed black lines are, respectively, the analytic results for regular/Poisson and exponential degree
      distributions (see Eqs. (\ref{kop1}-\ref{kop4})), while the symbols are numerical solutions of Eq.~(\ref{gh1}) with $c=100$.
      The numerical data for regular/Poisson graphs in panel (a) is a Gaussian distribution with variance of $O(1/c)$, approaching
      the Dirac delta distribution (vertical arrow) for $c \rightarrow \infty$.
}
\label{dense}
\end{center}
\end{figure}

If $p_{K^{\rm out}}$ is an exponential distribution, where $\alpha=2$, we obtain in the limit $c \rightarrow \infty$ \cite{SM}
\begin{align}
& p_R(r|\lambda_b) = \frac{2}{\pi} K_0 \left( 2 |r| \right)  \quad (\lambda_b\notin \mathbb{R}), \label{kop3} \\
& p_R(r|\lambda_{\rm isol}) = \sqrt{2}  \, \delta\left[ {\rm Im}( r)  \right]   \Theta\left[  {\rm Re} (r) \right]  e^{- \sqrt{2} {\rm Re} (r) }, \label{kop4}
\end{align}
where  $\Theta(x)$ is the Heaviside step function and $K_0(x)$ is a modified Bessel function of the second kind \cite{grad2007}.
Figure \ref{dense} illustrates the shape of the distributions $p_R$ given by Eqs.~(\ref{kop1}-\ref{kop4}), and compares them with numerical solutions of Eq. (\ref{gh1}) for $c=100$.   
The derivation of Eqs.~(\ref{kop1}-\ref{kop4}) is explained in the Supplemental Material \cite{SM}.

\paragraph{Conclusions.}
We have  shed light on the relationship between  graph topology and the localization of  right eigenvectors in directed random graphs. If the moments of the  outdegree distribution $p_{K^{\rm out}}$ are
finite, then
right eigenvectors at the edge of the spectrum are localized below a critical mean outdegree. It is striking that the critical points for the localization transitions are universal, in the
sense they only depend on the lower moments of the distribution $p_J$  of the edge weights, regardless of the network topology. Therefore, localization in directed random graphs is fundamentally different from localization in undirected graphs,  for which
degree fluctuations are important \cite{bauer2001random, krivelevich2003largest, Kuhn2008, Gol2012, metz2010localization, slanina2012,Nada2013,  Pastor2016}. Indeed, the eigenvector associated with the largest eigenvalue of the adjacency matrix of an 
 undirected random graph is  localized  if the maximal degree is large enough \cite{Gol2012}.    Degree-based localization is also possible for
directed random graphs, but then  $p_{K^{\rm out}}$  requires a divergent fourth moment.   

In the high connectivity limit, the distribution $p_R$  of the right eigenvector components is only
determined by the graph topology, independently of  $p_J$. If the outdegree fluctuations are small enough, then eigenvectors are delocalized and $p_R$
is given by the same universal distribution as in the case of Gaussian random matrices \cite{Porter1956,Mirlin2000}.   
On the other hand, if  the outdegree fluctuations are large enough, then eigenvectors are localized and the distribution $p_R$ depends on $p_{K^{\rm out}}$.
More generally, these results indicate that Gaussian random matrix theory describes well the spectral properties of high connectivity graphs only when
the degree fluctuations are sufficiently small \cite{MetzPRR}.
 
For future work, it would be interesting to explore the implications of eigenvector localization for the dynamics of neural networks  \cite{Amir2016, Zhang2019} and  ecosystems \cite{Allesina2015,Grilli2016}, to compare the theoretical predictions for the IPR with empirical values in real-world networks   \cite{dorogovtsev2003spectra, Pastor2018}, and to study eigenvector localization of Laplacians
of directed graphs  \cite{staring2003random, samukhin2008laplacian, kuhn2015spectra}.

\acknowledgements

The authors thank Jacopo Grilli for interesting discussions. F.L.M. thanks London Mathematical Laboratory and CNPq/Brazil for financial support. 

\bibliography{biblio}

\begin{thebibliography}{85}%
\makeatletter
\providecommand \@ifxundefined [1]{%
 \@ifx{#1\undefined}
}%
\providecommand \@ifnum [1]{%
 \ifnum #1\expandafter \@firstoftwo
 \else \expandafter \@secondoftwo
 \fi
}%
\providecommand \@ifx [1]{%
 \ifx #1\expandafter \@firstoftwo
 \else \expandafter \@secondoftwo
 \fi
}%
\providecommand \natexlab [1]{#1}%
\providecommand \enquote  [1]{``#1''}%
\providecommand \bibnamefont  [1]{#1}%
\providecommand \bibfnamefont [1]{#1}%
\providecommand \citenamefont [1]{#1}%
\providecommand \href@noop [0]{\@secondoftwo}%
\providecommand \href [0]{\begingroup \@sanitize@url \@href}%
\providecommand \@href[1]{\@@startlink{#1}\@@href}%
\providecommand \@@href[1]{\endgroup#1\@@endlink}%
\providecommand \@sanitize@url [0]{\catcode `\\12\catcode `\$12\catcode
  `\&12\catcode `\#12\catcode `\^12\catcode `\_12\catcode `\%12\relax}%
\providecommand \@@startlink[1]{}%
\providecommand \@@endlink[0]{}%
\providecommand \url  [0]{\begingroup\@sanitize@url \@url }%
\providecommand \@url [1]{\endgroup\@href {#1}{\urlprefix }}%
\providecommand \urlprefix  [0]{URL }%
\providecommand \Eprint [0]{\href }%
\providecommand \doibase [0]{http://dx.doi.org/}%
\providecommand \selectlanguage [0]{\@gobble}%
\providecommand \bibinfo  [0]{\@secondoftwo}%
\providecommand \bibfield  [0]{\@secondoftwo}%
\providecommand \translation [1]{[#1]}%
\providecommand \BibitemOpen [0]{}%
\providecommand \bibitemStop [0]{}%
\providecommand \bibitemNoStop [0]{.\EOS\space}%
\providecommand \EOS [0]{\spacefactor3000\relax}%
\providecommand \BibitemShut  [1]{\csname bibitem#1\endcsname}%
\let\auto@bib@innerbib\@empty
%</preamble>
\bibitem [{\citenamefont {Brunel}(2000)}]{brunel2000dynamics}%
  \BibitemOpen
  \bibfield  {author} {\bibinfo {author} {\bibfnamefont {Nicolas}\ \bibnamefont
  {Brunel}},\ }\bibfield  {title} {\enquote {\bibinfo {title} {Dynamics of
  sparsely connected networks of excitatory and inhibitory spiking neurons},}\
  }\href@noop {} {\bibfield  {journal} {\bibinfo  {journal} {Journal of
  computational neuroscience}\ }\textbf {\bibinfo {volume} {8}},\ \bibinfo
  {pages} {183--208} (\bibinfo {year} {2000})}\BibitemShut {NoStop}%
\bibitem [{\citenamefont {Bullmore}\ and\ \citenamefont
  {Sporns}(2009)}]{bullmore2009complex}%
  \BibitemOpen
  \bibfield  {author} {\bibinfo {author} {\bibfnamefont {Ed}~\bibnamefont
  {Bullmore}}\ and\ \bibinfo {author} {\bibfnamefont {Olaf}\ \bibnamefont
  {Sporns}},\ }\bibfield  {title} {\enquote {\bibinfo {title} {Complex brain
  networks: graph theoretical analysis of structural and functional systems},}\
  }\href@noop {} {\bibfield  {journal} {\bibinfo  {journal} {Nature reviews
  neuroscience}\ }\textbf {\bibinfo {volume} {10}},\ \bibinfo {pages}
  {186--198} (\bibinfo {year} {2009})}\BibitemShut {NoStop}%
\bibitem [{\citenamefont {Sporns}(2010)}]{sporns2010networks}%
  \BibitemOpen
  \bibfield  {author} {\bibinfo {author} {\bibfnamefont {Olaf}\ \bibnamefont
  {Sporns}},\ }\href@noop {} {\emph {\bibinfo {title} {Networks of the
  Brain}}}\ (\bibinfo  {publisher} {MIT press},\ \bibinfo {year}
  {2010})\BibitemShut {NoStop}%
\bibitem [{\citenamefont {Bascompte}(2009)}]{bascompte2009disentangling}%
  \BibitemOpen
  \bibfield  {author} {\bibinfo {author} {\bibfnamefont {Jordi}\ \bibnamefont
  {Bascompte}},\ }\bibfield  {title} {\enquote {\bibinfo {title} {Disentangling
  the web of life},}\ }\href@noop {} {\bibfield  {journal} {\bibinfo  {journal}
  {Science}\ }\textbf {\bibinfo {volume} {325}},\ \bibinfo {pages} {416--419}
  (\bibinfo {year} {2009})}\BibitemShut {NoStop}%
\bibitem [{\citenamefont {Milo}\ \emph {et~al.}(2002)\citenamefont {Milo},
  \citenamefont {Shen-Orr}, \citenamefont {Itzkovitz}, \citenamefont {Kashtan},
  \citenamefont {Chklovskii},\ and\ \citenamefont {Alon}}]{milo2002network}%
  \BibitemOpen
  \bibfield  {author} {\bibinfo {author} {\bibfnamefont {Ron}\ \bibnamefont
  {Milo}}, \bibinfo {author} {\bibfnamefont {Shai}\ \bibnamefont {Shen-Orr}},
  \bibinfo {author} {\bibfnamefont {Shalev}\ \bibnamefont {Itzkovitz}},
  \bibinfo {author} {\bibfnamefont {Nadav}\ \bibnamefont {Kashtan}}, \bibinfo
  {author} {\bibfnamefont {Dmitri}\ \bibnamefont {Chklovskii}}, \ and\ \bibinfo
  {author} {\bibfnamefont {Uri}\ \bibnamefont {Alon}},\ }\bibfield  {title}
  {\enquote {\bibinfo {title} {Network motifs: simple building blocks of
  complex networks},}\ }\href@noop {} {\bibfield  {journal} {\bibinfo
  {journal} {Science}\ }\textbf {\bibinfo {volume} {298}},\ \bibinfo {pages}
  {824--827} (\bibinfo {year} {2002})}\BibitemShut {NoStop}%
\bibitem [{\citenamefont {Shen-Orr}\ \emph {et~al.}(2002)\citenamefont
  {Shen-Orr}, \citenamefont {Milo}, \citenamefont {Mangan},\ and\ \citenamefont
  {Alon}}]{shen2002network}%
  \BibitemOpen
  \bibfield  {author} {\bibinfo {author} {\bibfnamefont {Shai~S}\ \bibnamefont
  {Shen-Orr}}, \bibinfo {author} {\bibfnamefont {Ron}\ \bibnamefont {Milo}},
  \bibinfo {author} {\bibfnamefont {Shmoolik}\ \bibnamefont {Mangan}}, \ and\
  \bibinfo {author} {\bibfnamefont {Uri}\ \bibnamefont {Alon}},\ }\bibfield
  {title} {\enquote {\bibinfo {title} {Network motifs in the transcriptional
  regulation network of escherichia coli},}\ }\href@noop {} {\bibfield
  {journal} {\bibinfo  {journal} {Nature genetics}\ }\textbf {\bibinfo {volume}
  {31}},\ \bibinfo {pages} {64--68} (\bibinfo {year} {2002})}\BibitemShut
  {NoStop}%
\bibitem [{\citenamefont {Lee}\ and\ \citenamefont
  {et~al}(2002)}]{lee2002transcriptional}%
  \BibitemOpen
  \bibfield  {author} {\bibinfo {author} {\bibfnamefont {T~I}\ \bibnamefont
  {Lee}}\ and\ \bibinfo {author} {\bibnamefont {et~al}},\ }\bibfield  {title}
  {\enquote {\bibinfo {title} {Transcriptional regulatory networks in
  saccharomyces cerevisiae},}\ }\href@noop {} {\bibfield  {journal} {\bibinfo
  {journal} {science}\ }\textbf {\bibinfo {volume} {298}},\ \bibinfo {pages}
  {799--804} (\bibinfo {year} {2002})}\BibitemShut {NoStop}%
\bibitem [{\citenamefont {Kwak}\ \emph {et~al.}(2010)\citenamefont {Kwak},
  \citenamefont {Lee}, \citenamefont {Park},\ and\ \citenamefont
  {Moon}}]{kwak2010twitter}%
  \BibitemOpen
  \bibfield  {author} {\bibinfo {author} {\bibfnamefont {Haewoon}\ \bibnamefont
  {Kwak}}, \bibinfo {author} {\bibfnamefont {Changhyun}\ \bibnamefont {Lee}},
  \bibinfo {author} {\bibfnamefont {Hosung}\ \bibnamefont {Park}}, \ and\
  \bibinfo {author} {\bibfnamefont {Sue}\ \bibnamefont {Moon}},\ }\bibfield
  {title} {\enquote {\bibinfo {title} {What is twitter, a social network or a
  news media?}}\ }in\ \href@noop {} {\emph {\bibinfo {booktitle} {Proceedings
  of the 19th international conference on World wide web}}}\ (\bibinfo {year}
  {2010})\ pp.\ \bibinfo {pages} {591--600}\BibitemShut {NoStop}%
\bibitem [{\citenamefont {Aiello}\ \emph {et~al.}(2012)\citenamefont {Aiello},
  \citenamefont {Barrat}, \citenamefont {Schifanella}, \citenamefont {Cattuto},
  \citenamefont {Markines},\ and\ \citenamefont
  {Menczer}}]{aiello2012friendship}%
  \BibitemOpen
  \bibfield  {author} {\bibinfo {author} {\bibfnamefont {Luca~Maria}\
  \bibnamefont {Aiello}}, \bibinfo {author} {\bibfnamefont {Alain}\
  \bibnamefont {Barrat}}, \bibinfo {author} {\bibfnamefont {Rossano}\
  \bibnamefont {Schifanella}}, \bibinfo {author} {\bibfnamefont {Ciro}\
  \bibnamefont {Cattuto}}, \bibinfo {author} {\bibfnamefont {Benjamin}\
  \bibnamefont {Markines}}, \ and\ \bibinfo {author} {\bibfnamefont {Filippo}\
  \bibnamefont {Menczer}},\ }\bibfield  {title} {\enquote {\bibinfo {title}
  {Friendship prediction and homophily in social media},}\ }\href@noop {}
  {\bibfield  {journal} {\bibinfo  {journal} {ACM Transactions on the Web
  (TWEB)}\ }\textbf {\bibinfo {volume} {6}},\ \bibinfo {pages} {1--33}
  (\bibinfo {year} {2012})}\BibitemShut {NoStop}%
\bibitem [{\citenamefont {Broder}\ \emph {et~al.}(2000)\citenamefont {Broder},
  \citenamefont {Kumar}, \citenamefont {Maghoul}, \citenamefont {Raghavan},
  \citenamefont {Rajagopalan}, \citenamefont {Stata}, \citenamefont {Tomkins},\
  and\ \citenamefont {Wiener}}]{broder2000graph}%
  \BibitemOpen
  \bibfield  {author} {\bibinfo {author} {\bibfnamefont {Andrei}\ \bibnamefont
  {Broder}}, \bibinfo {author} {\bibfnamefont {Ravi}\ \bibnamefont {Kumar}},
  \bibinfo {author} {\bibfnamefont {Farzin}\ \bibnamefont {Maghoul}}, \bibinfo
  {author} {\bibfnamefont {Prabhakar}\ \bibnamefont {Raghavan}}, \bibinfo
  {author} {\bibfnamefont {Sridhar}\ \bibnamefont {Rajagopalan}}, \bibinfo
  {author} {\bibfnamefont {Raymie}\ \bibnamefont {Stata}}, \bibinfo {author}
  {\bibfnamefont {Andrew}\ \bibnamefont {Tomkins}}, \ and\ \bibinfo {author}
  {\bibfnamefont {Janet}\ \bibnamefont {Wiener}},\ }\bibfield  {title}
  {\enquote {\bibinfo {title} {Graph structure in the web},}\ }\href@noop {}
  {\bibfield  {journal} {\bibinfo  {journal} {Computer networks}\ }\textbf
  {\bibinfo {volume} {33}},\ \bibinfo {pages} {309--320} (\bibinfo {year}
  {2000})}\BibitemShut {NoStop}%
\bibitem [{\citenamefont {Pastor-Satorras}\ and\ \citenamefont
  {Vespignani}(2007)}]{pastor2007evolution}%
  \BibitemOpen
  \bibfield  {author} {\bibinfo {author} {\bibfnamefont {Romualdo}\
  \bibnamefont {Pastor-Satorras}}\ and\ \bibinfo {author} {\bibfnamefont
  {Alessandro}\ \bibnamefont {Vespignani}},\ }\href@noop {} {\emph {\bibinfo
  {title} {Evolution and structure of the Internet: A statistical physics
  approach}}}\ (\bibinfo  {publisher} {Cambridge University Press},\ \bibinfo
  {year} {2007})\BibitemShut {NoStop}%
\bibitem [{\citenamefont {Hartman}(1960)}]{hartman1960lemma}%
  \BibitemOpen
  \bibfield  {author} {\bibinfo {author} {\bibfnamefont {Philip}\ \bibnamefont
  {Hartman}},\ }\bibfield  {title} {\enquote {\bibinfo {title} {A lemma in the
  theory of structural stability of differential equations},}\ }\href@noop {}
  {\bibfield  {journal} {\bibinfo  {journal} {Proceedings of the American
  Mathematical Society}\ }\textbf {\bibinfo {volume} {11}},\ \bibinfo {pages}
  {610--620} (\bibinfo {year} {1960})}\BibitemShut {NoStop}%
\bibitem [{\citenamefont {Grobman}(1959)}]{grobman1959homeomorphism}%
  \BibitemOpen
  \bibfield  {author} {\bibinfo {author} {\bibfnamefont {David~M}\ \bibnamefont
  {Grobman}},\ }\bibfield  {title} {\enquote {\bibinfo {title} {Homeomorphism
  of systems of differential equations},}\ }\href@noop {} {\bibfield  {journal}
  {\bibinfo  {journal} {Doklady Akademii Nauk SSSR}\ }\textbf {\bibinfo
  {volume} {128}},\ \bibinfo {pages} {880--881} (\bibinfo {year}
  {1959})}\BibitemShut {NoStop}%
\bibitem [{\citenamefont {Sompolinsky}\ \emph {et~al.}(1988)\citenamefont
  {Sompolinsky}, \citenamefont {Crisanti},\ and\ \citenamefont
  {Sommers}}]{sompolinsky1988chaos}%
  \BibitemOpen
  \bibfield  {author} {\bibinfo {author} {\bibfnamefont {Haim}\ \bibnamefont
  {Sompolinsky}}, \bibinfo {author} {\bibfnamefont {Andrea}\ \bibnamefont
  {Crisanti}}, \ and\ \bibinfo {author} {\bibfnamefont {Hans-Jurgen}\
  \bibnamefont {Sommers}},\ }\bibfield  {title} {\enquote {\bibinfo {title}
  {Chaos in random neural networks},}\ }\href@noop {} {\bibfield  {journal}
  {\bibinfo  {journal} {Physical review letters}\ }\textbf {\bibinfo {volume}
  {61}},\ \bibinfo {pages} {259} (\bibinfo {year} {1988})}\BibitemShut
  {NoStop}%
\bibitem [{\citenamefont {Del~Molino}\ \emph {et~al.}(2013)\citenamefont
  {Del~Molino}, \citenamefont {Pakdaman}, \citenamefont {Touboul},\ and\
  \citenamefont {Wainrib}}]{del2013synchronization}%
  \BibitemOpen
  \bibfield  {author} {\bibinfo {author} {\bibfnamefont {Luis
  Carlos~Garc{\'\i}a}\ \bibnamefont {Del~Molino}}, \bibinfo {author}
  {\bibfnamefont {Khashayar}\ \bibnamefont {Pakdaman}}, \bibinfo {author}
  {\bibfnamefont {Jonathan}\ \bibnamefont {Touboul}}, \ and\ \bibinfo {author}
  {\bibfnamefont {Gilles}\ \bibnamefont {Wainrib}},\ }\bibfield  {title}
  {\enquote {\bibinfo {title} {Synchronization in random balanced networks},}\
  }\href@noop {} {\bibfield  {journal} {\bibinfo  {journal} {Physical Review
  E}\ }\textbf {\bibinfo {volume} {88}},\ \bibinfo {pages} {042824} (\bibinfo
  {year} {2013})}\BibitemShut {NoStop}%
\bibitem [{\citenamefont {Kadmon}\ and\ \citenamefont
  {Sompolinsky}(2015)}]{kadmon2015transition}%
  \BibitemOpen
  \bibfield  {author} {\bibinfo {author} {\bibfnamefont {Jonathan}\
  \bibnamefont {Kadmon}}\ and\ \bibinfo {author} {\bibfnamefont {Haim}\
  \bibnamefont {Sompolinsky}},\ }\bibfield  {title} {\enquote {\bibinfo {title}
  {Transition to chaos in random neuronal networks},}\ }\href@noop {}
  {\bibfield  {journal} {\bibinfo  {journal} {Physical Review X}\ }\textbf
  {\bibinfo {volume} {5}},\ \bibinfo {pages} {041030} (\bibinfo {year}
  {2015})}\BibitemShut {NoStop}%
\bibitem [{\citenamefont {Aljadeff}\ \emph {et~al.}(2015)\citenamefont
  {Aljadeff}, \citenamefont {Stern},\ and\ \citenamefont
  {Sharpee}}]{aljadeff2015transition}%
  \BibitemOpen
  \bibfield  {author} {\bibinfo {author} {\bibfnamefont {Johnatan}\
  \bibnamefont {Aljadeff}}, \bibinfo {author} {\bibfnamefont {Merav}\
  \bibnamefont {Stern}}, \ and\ \bibinfo {author} {\bibfnamefont {Tatyana}\
  \bibnamefont {Sharpee}},\ }\bibfield  {title} {\enquote {\bibinfo {title}
  {Transition to chaos in random networks with cell-type-specific
  connectivity},}\ }\href@noop {} {\bibfield  {journal} {\bibinfo  {journal}
  {Physical review letters}\ }\textbf {\bibinfo {volume} {114}},\ \bibinfo
  {pages} {088101} (\bibinfo {year} {2015})}\BibitemShut {NoStop}%
\bibitem [{\citenamefont {Mart{\'\i}}\ \emph {et~al.}(2018)\citenamefont
  {Mart{\'\i}}, \citenamefont {Brunel},\ and\ \citenamefont
  {Ostojic}}]{marti2018correlations}%
  \BibitemOpen
  \bibfield  {author} {\bibinfo {author} {\bibfnamefont {Daniel}\ \bibnamefont
  {Mart{\'\i}}}, \bibinfo {author} {\bibfnamefont {Nicolas}\ \bibnamefont
  {Brunel}}, \ and\ \bibinfo {author} {\bibfnamefont {Srdjan}\ \bibnamefont
  {Ostojic}},\ }\bibfield  {title} {\enquote {\bibinfo {title} {Correlations
  between synapses in pairs of neurons slow down dynamics in randomly connected
  neural networks},}\ }\href@noop {} {\bibfield  {journal} {\bibinfo  {journal}
  {Physical Review E}\ }\textbf {\bibinfo {volume} {97}},\ \bibinfo {pages}
  {062314} (\bibinfo {year} {2018})}\BibitemShut {NoStop}%
\bibitem [{\citenamefont {May}(1972)}]{May}%
  \BibitemOpen
  \bibfield  {author} {\bibinfo {author} {\bibfnamefont {Robert~M.}\
  \bibnamefont {May}},\ }\bibfield  {title} {\enquote {\bibinfo {title} {Will a
  large complex system be stable?}}\ }\href@noop {} {\bibfield  {journal}
  {\bibinfo  {journal} {Nature}\ }\textbf {\bibinfo {volume} {238}},\ \bibinfo
  {pages} {413–414} (\bibinfo {year} {1972})}\BibitemShut {NoStop}%
\bibitem [{\citenamefont {Allesina}\ \emph {et~al.}(2015)\citenamefont
  {Allesina}, \citenamefont {Grilli}, \citenamefont {Barab\'as}, \citenamefont
  {Tang},\ and\ \citenamefont {Aljadeff}}]{Allesina2015}%
  \BibitemOpen
  \bibfield  {author} {\bibinfo {author} {\bibfnamefont {Stefano}\ \bibnamefont
  {Allesina}}, \bibinfo {author} {\bibfnamefont {Jacopo}\ \bibnamefont
  {Grilli}}, \bibinfo {author} {\bibfnamefont {Gy\"orgy}\ \bibnamefont
  {Barab\'as}}, \bibinfo {author} {\bibfnamefont {Si}~\bibnamefont {Tang}}, \
  and\ \bibinfo {author} {\bibfnamefont {Johnatan}\ \bibnamefont {Aljadeff}},\
  }\bibfield  {title} {\enquote {\bibinfo {title} {Predicting the stability of
  large structured food webs},}\ }\href@noop {} {\bibfield  {journal} {\bibinfo
   {journal} {Nat. Commun.}\ }\textbf {\bibinfo {volume} {6}},\ \bibinfo
  {pages} {7842} (\bibinfo {year} {2015})}\BibitemShut {NoStop}%
\bibitem [{\citenamefont {Gibbs}\ \emph {et~al.}(2018)\citenamefont {Gibbs},
  \citenamefont {Grilli}, \citenamefont {Rogers},\ and\ \citenamefont
  {Allesina}}]{gibbs2018effect}%
  \BibitemOpen
  \bibfield  {author} {\bibinfo {author} {\bibfnamefont {Theo}\ \bibnamefont
  {Gibbs}}, \bibinfo {author} {\bibfnamefont {Jacopo}\ \bibnamefont {Grilli}},
  \bibinfo {author} {\bibfnamefont {Tim}\ \bibnamefont {Rogers}}, \ and\
  \bibinfo {author} {\bibfnamefont {Stefano}\ \bibnamefont {Allesina}},\
  }\bibfield  {title} {\enquote {\bibinfo {title} {Effect of population
  abundances on the stability of large random ecosystems},}\ }\href@noop {}
  {\bibfield  {journal} {\bibinfo  {journal} {Physical Review E}\ }\textbf
  {\bibinfo {volume} {98}},\ \bibinfo {pages} {022410} (\bibinfo {year}
  {2018})}\BibitemShut {NoStop}%
\bibitem [{\citenamefont {Chen}\ \emph {et~al.}(2019)\citenamefont {Chen},
  \citenamefont {Shen}, \citenamefont {Lin}, \citenamefont {Tong},
  \citenamefont {Zhao}, \citenamefont {Allesina}, \citenamefont {Shen},\ and\
  \citenamefont {Wu}}]{Chen2019}%
  \BibitemOpen
  \bibfield  {author} {\bibinfo {author} {\bibfnamefont {Yuxin}\ \bibnamefont
  {Chen}}, \bibinfo {author} {\bibfnamefont {Yang}\ \bibnamefont {Shen}},
  \bibinfo {author} {\bibfnamefont {Pei}\ \bibnamefont {Lin}}, \bibinfo
  {author} {\bibfnamefont {Ding}\ \bibnamefont {Tong}}, \bibinfo {author}
  {\bibfnamefont {Yixin}\ \bibnamefont {Zhao}}, \bibinfo {author}
  {\bibfnamefont {Stefano}\ \bibnamefont {Allesina}}, \bibinfo {author}
  {\bibfnamefont {Xu}~\bibnamefont {Shen}}, \ and\ \bibinfo {author}
  {\bibfnamefont {Chung-I}\ \bibnamefont {Wu}},\ }\bibfield  {title} {\enquote
  {\bibinfo {title} {{Gene regulatory network stabilized by pervasive weak
  repressions: microRNA functions revealed by the May–Wigner theory}},}\
  }\href {\doibase 10.1093/nsr/nwz076} {\bibfield  {journal} {\bibinfo
  {journal} {National Science Review}\ }\textbf {\bibinfo {volume} {6}},\
  \bibinfo {pages} {1176--1188} (\bibinfo {year} {2019})},\ \Eprint
  {http://arxiv.org/abs/https://academic.oup.com/nsr/article-pdf/6/6/1176/32351125/nwz076.pdf}
  {https://academic.oup.com/nsr/article-pdf/6/6/1176/32351125/nwz076.pdf}
  \BibitemShut {NoStop}%
\bibitem [{\citenamefont {Guo}\ and\ \citenamefont {Amir}(2020)}]{Guo2020}%
  \BibitemOpen
  \bibfield  {author} {\bibinfo {author} {\bibfnamefont {Yipei}\ \bibnamefont
  {Guo}}\ and\ \bibinfo {author} {\bibfnamefont {Ariel}\ \bibnamefont {Amir}},\
  }\href@noop {} {\enquote {\bibinfo {title} {Stability of gene regulatory
  networks},}\ } (\bibinfo {year} {2020}),\ \Eprint
  {http://arxiv.org/abs/2006.00018} {arXiv:2006.00018 [physics.bio-ph]}
  \BibitemShut {NoStop}%
\bibitem [{\citenamefont {Mieghem}(2012)}]{Van2012}%
  \BibitemOpen
  \bibfield  {author} {\bibinfo {author} {\bibfnamefont {P.~Van}\ \bibnamefont
  {Mieghem}},\ }\bibfield  {title} {\enquote {\bibinfo {title} {Epidemic phase
  transition of the {SIS} type in networks},}\ }\href {\doibase
  10.1209/0295-5075/97/48004} {\bibfield  {journal} {\bibinfo  {journal} {{EPL}
  (Europhysics Letters)}\ }\textbf {\bibinfo {volume} {97}},\ \bibinfo {pages}
  {48004} (\bibinfo {year} {2012})}\BibitemShut {NoStop}%
\bibitem [{\citenamefont {Goltsev}\ \emph {et~al.}(2012)\citenamefont
  {Goltsev}, \citenamefont {Dorogovtsev}, \citenamefont {Oliveira},\ and\
  \citenamefont {Mendes}}]{Gol2012}%
  \BibitemOpen
  \bibfield  {author} {\bibinfo {author} {\bibfnamefont {A.~V.}\ \bibnamefont
  {Goltsev}}, \bibinfo {author} {\bibfnamefont {S.~N.}\ \bibnamefont
  {Dorogovtsev}}, \bibinfo {author} {\bibfnamefont {J.~G.}\ \bibnamefont
  {Oliveira}}, \ and\ \bibinfo {author} {\bibfnamefont {J.~F.~F.}\ \bibnamefont
  {Mendes}},\ }\bibfield  {title} {\enquote {\bibinfo {title} {Localization and
  spreading of diseases in complex networks},}\ }\href {\doibase
  10.1103/PhysRevLett.109.128702} {\bibfield  {journal} {\bibinfo  {journal}
  {Phys. Rev. Lett.}\ }\textbf {\bibinfo {volume} {109}},\ \bibinfo {pages}
  {128702} (\bibinfo {year} {2012})}\BibitemShut {NoStop}%
\bibitem [{\citenamefont {Pastor-Satorras}\ and\ \citenamefont
  {Castellano}(2018)}]{Pastor2018}%
  \BibitemOpen
  \bibfield  {author} {\bibinfo {author} {\bibfnamefont {R.}~\bibnamefont
  {Pastor-Satorras}}\ and\ \bibinfo {author} {\bibfnamefont {C.}~\bibnamefont
  {Castellano}},\ }\bibfield  {title} {\enquote {\bibinfo {title} {Eigenvector
  localization in real networks and its implications for epidemic spreading},}\
  }\href@noop {} {\bibfield  {journal} {\bibinfo  {journal} {J. Stat. Phys.}\
  }\textbf {\bibinfo {volume} {173}},\ \bibinfo {pages} {1110--1123} (\bibinfo
  {year} {2018})}\BibitemShut {NoStop}%
\bibitem [{\citenamefont {Li}\ \emph {et~al.}(2013)\citenamefont {Li},
  \citenamefont {Wang},\ and\ \citenamefont {Van~Mieghem}}]{Li2013}%
  \BibitemOpen
  \bibfield  {author} {\bibinfo {author} {\bibfnamefont {Cong}\ \bibnamefont
  {Li}}, \bibinfo {author} {\bibfnamefont {Huijuan}\ \bibnamefont {Wang}}, \
  and\ \bibinfo {author} {\bibfnamefont {Piet}\ \bibnamefont {Van~Mieghem}},\
  }\bibfield  {title} {\enquote {\bibinfo {title} {Epidemic threshold in
  directed networks},}\ }\href {\doibase 10.1103/PhysRevE.88.062802} {\bibfield
   {journal} {\bibinfo  {journal} {Phys. Rev. E}\ }\textbf {\bibinfo {volume}
  {88}},\ \bibinfo {pages} {062802} (\bibinfo {year} {2013})}\BibitemShut
  {NoStop}%
\bibitem [{\citenamefont {Silva}\ and\ \citenamefont
  {Ferreira}(2020)}]{silva2020dissecting}%
  \BibitemOpen
  \bibfield  {author} {\bibinfo {author} {\bibfnamefont {Diogo~H}\ \bibnamefont
  {Silva}}\ and\ \bibinfo {author} {\bibfnamefont {Silvio~C}\ \bibnamefont
  {Ferreira}},\ }\bibfield  {title} {\enquote {\bibinfo {title} {Dissecting
  localization phenomena of dynamical processes on networks},}\ }\href@noop {}
  {\bibfield  {journal} {\bibinfo  {journal} {arXiv preprint arXiv:2011.10918}\
  } (\bibinfo {year} {2020})}\BibitemShut {NoStop}%
\bibitem [{\citenamefont {Bonacich}(1972)}]{Bona1972}%
  \BibitemOpen
  \bibfield  {author} {\bibinfo {author} {\bibfnamefont {Phillip}\ \bibnamefont
  {Bonacich}},\ }\bibfield  {title} {\enquote {\bibinfo {title} {Factoring and
  weighting approaches to status scores and clique identification},}\ }\href
  {\doibase 10.1080/0022250X.1972.9989806} {\bibfield  {journal} {\bibinfo
  {journal} {The Journal of Mathematical Sociology}\ }\textbf {\bibinfo
  {volume} {2}},\ \bibinfo {pages} {113--120} (\bibinfo {year} {1972})},\
  \Eprint {http://arxiv.org/abs/https://doi.org/10.1080/0022250X.1972.9989806}
  {https://doi.org/10.1080/0022250X.1972.9989806} \BibitemShut {NoStop}%
\bibitem [{\citenamefont {Restrepo}\ \emph {et~al.}(2006)\citenamefont
  {Restrepo}, \citenamefont {Ott},\ and\ \citenamefont {Hunt}}]{Rest2006a}%
  \BibitemOpen
  \bibfield  {author} {\bibinfo {author} {\bibfnamefont {Juan~G.}\ \bibnamefont
  {Restrepo}}, \bibinfo {author} {\bibfnamefont {Edward}\ \bibnamefont {Ott}},
  \ and\ \bibinfo {author} {\bibfnamefont {Brian~R.}\ \bibnamefont {Hunt}},\
  }\bibfield  {title} {\enquote {\bibinfo {title} {Characterizing the dynamical
  importance of network nodes and links},}\ }\href {\doibase
  10.1103/PhysRevLett.97.094102} {\bibfield  {journal} {\bibinfo  {journal}
  {Phys. Rev. Lett.}\ }\textbf {\bibinfo {volume} {97}},\ \bibinfo {pages}
  {094102} (\bibinfo {year} {2006})}\BibitemShut {NoStop}%
\bibitem [{\citenamefont {Martin}\ \emph {et~al.}(2014)\citenamefont {Martin},
  \citenamefont {Zhang},\ and\ \citenamefont {Newman}}]{Travis2014}%
  \BibitemOpen
  \bibfield  {author} {\bibinfo {author} {\bibfnamefont {Travis}\ \bibnamefont
  {Martin}}, \bibinfo {author} {\bibfnamefont {Xiao}\ \bibnamefont {Zhang}}, \
  and\ \bibinfo {author} {\bibfnamefont {M.~E.~J.}\ \bibnamefont {Newman}},\
  }\bibfield  {title} {\enquote {\bibinfo {title} {Localization and centrality
  in networks},}\ }\href {\doibase 10.1103/PhysRevE.90.052808} {\bibfield
  {journal} {\bibinfo  {journal} {Phys. Rev. E}\ }\textbf {\bibinfo {volume}
  {90}},\ \bibinfo {pages} {052808} (\bibinfo {year} {2014})}\BibitemShut
  {NoStop}%
\bibitem [{\citenamefont {Krzakala}\ \emph {et~al.}(2013)\citenamefont
  {Krzakala}, \citenamefont {Moore}, \citenamefont {Mossel}, \citenamefont
  {Neeman}, \citenamefont {Sly}, \citenamefont {Zdeborov{\'a}},\ and\
  \citenamefont {Zhang}}]{krzakala2013spectral}%
  \BibitemOpen
  \bibfield  {author} {\bibinfo {author} {\bibfnamefont {Florent}\ \bibnamefont
  {Krzakala}}, \bibinfo {author} {\bibfnamefont {Cristopher}\ \bibnamefont
  {Moore}}, \bibinfo {author} {\bibfnamefont {Elchanan}\ \bibnamefont
  {Mossel}}, \bibinfo {author} {\bibfnamefont {Joe}\ \bibnamefont {Neeman}},
  \bibinfo {author} {\bibfnamefont {Allan}\ \bibnamefont {Sly}}, \bibinfo
  {author} {\bibfnamefont {Lenka}\ \bibnamefont {Zdeborov{\'a}}}, \ and\
  \bibinfo {author} {\bibfnamefont {Pan}\ \bibnamefont {Zhang}},\ }\bibfield
  {title} {\enquote {\bibinfo {title} {Spectral redemption in clustering sparse
  networks},}\ }\href@noop {} {\bibfield  {journal} {\bibinfo  {journal}
  {Proceedings of the National Academy of Sciences}\ }\textbf {\bibinfo
  {volume} {110}},\ \bibinfo {pages} {20935--20940} (\bibinfo {year}
  {2013})}\BibitemShut {NoStop}%
\bibitem [{\citenamefont {Bordenave}\ \emph {et~al.}(2015)\citenamefont
  {Bordenave}, \citenamefont {Lelarge},\ and\ \citenamefont
  {Massouli{\'e}}}]{bordenave2015non}%
  \BibitemOpen
  \bibfield  {author} {\bibinfo {author} {\bibfnamefont {Charles}\ \bibnamefont
  {Bordenave}}, \bibinfo {author} {\bibfnamefont {Marc}\ \bibnamefont
  {Lelarge}}, \ and\ \bibinfo {author} {\bibfnamefont {Laurent}\ \bibnamefont
  {Massouli{\'e}}},\ }\bibfield  {title} {\enquote {\bibinfo {title}
  {Non-backtracking spectrum of random graphs: community detection and
  non-regular ramanujan graphs},}\ }in\ \href@noop {} {\emph {\bibinfo
  {booktitle} {2015 IEEE 56th Annual Symposium on Foundations of Computer
  Science}}}\ (\bibinfo {organization} {IEEE},\ \bibinfo {year} {2015})\ pp.\
  \bibinfo {pages} {1347--1357}\BibitemShut {NoStop}%
\bibitem [{\citenamefont {Kawamoto}(2018)}]{kawamoto2018algorithmic}%
  \BibitemOpen
  \bibfield  {author} {\bibinfo {author} {\bibfnamefont {Tatsuro}\ \bibnamefont
  {Kawamoto}},\ }\bibfield  {title} {\enquote {\bibinfo {title} {Algorithmic
  detectability threshold of the stochastic block model},}\ }\href@noop {}
  {\bibfield  {journal} {\bibinfo  {journal} {Physical Review E}\ }\textbf
  {\bibinfo {volume} {97}},\ \bibinfo {pages} {032301} (\bibinfo {year}
  {2018})}\BibitemShut {NoStop}%
\bibitem [{\citenamefont {Bordenave}\ \emph {et~al.}(2020)\citenamefont
  {Bordenave}, \citenamefont {Coste},\ and\ \citenamefont
  {Nadakuditi}}]{bordenave2020detection}%
  \BibitemOpen
  \bibfield  {author} {\bibinfo {author} {\bibfnamefont {Charles}\ \bibnamefont
  {Bordenave}}, \bibinfo {author} {\bibfnamefont {Simon}\ \bibnamefont
  {Coste}}, \ and\ \bibinfo {author} {\bibfnamefont {Raj~Rao}\ \bibnamefont
  {Nadakuditi}},\ }\bibfield  {title} {\enquote {\bibinfo {title} {Detection
  thresholds in very sparse matrix completion},}\ }\href@noop {} {\bibfield
  {journal} {\bibinfo  {journal} {arXiv preprint arXiv:2005.06062}\ } (\bibinfo
  {year} {2020})}\BibitemShut {NoStop}%
\bibitem [{\citenamefont {Abou-Chacra}\ \emph {et~al.}(1973)\citenamefont
  {Abou-Chacra}, \citenamefont {Thouless},\ and\ \citenamefont
  {Anderson}}]{abou1973selfconsistent}%
  \BibitemOpen
  \bibfield  {author} {\bibinfo {author} {\bibfnamefont {Ragi}\ \bibnamefont
  {Abou-Chacra}}, \bibinfo {author} {\bibfnamefont {DJ}~\bibnamefont
  {Thouless}}, \ and\ \bibinfo {author} {\bibfnamefont {PW}~\bibnamefont
  {Anderson}},\ }\bibfield  {title} {\enquote {\bibinfo {title} {A
  selfconsistent theory of localization},}\ }\href@noop {} {\bibfield
  {journal} {\bibinfo  {journal} {Journal of Physics C: Solid State Physics}\
  }\textbf {\bibinfo {volume} {6}},\ \bibinfo {pages} {1734} (\bibinfo {year}
  {1973})}\BibitemShut {NoStop}%
\bibitem [{\citenamefont {Aizenman}\ and\ \citenamefont
  {Warzel}(2011)}]{aizenman2011extended}%
  \BibitemOpen
  \bibfield  {author} {\bibinfo {author} {\bibfnamefont {Michael}\ \bibnamefont
  {Aizenman}}\ and\ \bibinfo {author} {\bibfnamefont {Simone}\ \bibnamefont
  {Warzel}},\ }\bibfield  {title} {\enquote {\bibinfo {title} {Extended states
  in a lifshitz tail regime for random schr{\"o}dinger operators on trees},}\
  }\href@noop {} {\bibfield  {journal} {\bibinfo  {journal} {Physical review
  letters}\ }\textbf {\bibinfo {volume} {106}},\ \bibinfo {pages} {136804}
  (\bibinfo {year} {2011})}\BibitemShut {NoStop}%
\bibitem [{\citenamefont {Giraud}\ \emph {et~al.}(2009)\citenamefont {Giraud},
  \citenamefont {Georgeot},\ and\ \citenamefont
  {Shepelyansky}}]{PhysRevE.80.026107}%
  \BibitemOpen
  \bibfield  {author} {\bibinfo {author} {\bibfnamefont {Olivier}\ \bibnamefont
  {Giraud}}, \bibinfo {author} {\bibfnamefont {Bertrand}\ \bibnamefont
  {Georgeot}}, \ and\ \bibinfo {author} {\bibfnamefont {Dima~L.}\ \bibnamefont
  {Shepelyansky}},\ }\bibfield  {title} {\enquote {\bibinfo {title}
  {Delocalization transition for the google matrix},}\ }\href {\doibase
  10.1103/PhysRevE.80.026107} {\bibfield  {journal} {\bibinfo  {journal} {Phys.
  Rev. E}\ }\textbf {\bibinfo {volume} {80}},\ \bibinfo {pages} {026107}
  (\bibinfo {year} {2009})}\BibitemShut {NoStop}%
\bibitem [{\citenamefont {Fyodorov}\ and\ \citenamefont
  {Mirlin}(1991)}]{Fyodorov1991}%
  \BibitemOpen
  \bibfield  {author} {\bibinfo {author} {\bibfnamefont {Yan~V.}\ \bibnamefont
  {Fyodorov}}\ and\ \bibinfo {author} {\bibfnamefont {Alexander~D.}\
  \bibnamefont {Mirlin}},\ }\bibfield  {title} {\enquote {\bibinfo {title}
  {Localization in ensemble of sparse random matrices},}\ }\href {\doibase
  10.1103/PhysRevLett.67.2049} {\bibfield  {journal} {\bibinfo  {journal}
  {Phys. Rev. Lett.}\ }\textbf {\bibinfo {volume} {67}},\ \bibinfo {pages}
  {2049--2052} (\bibinfo {year} {1991})}\BibitemShut {NoStop}%
\bibitem [{\citenamefont {Mirlin}\ and\ \citenamefont
  {Fyodorov}(1991)}]{mirlin1991localization}%
  \BibitemOpen
  \bibfield  {author} {\bibinfo {author} {\bibfnamefont {Alexander~D}\
  \bibnamefont {Mirlin}}\ and\ \bibinfo {author} {\bibfnamefont {Yan~V}\
  \bibnamefont {Fyodorov}},\ }\bibfield  {title} {\enquote {\bibinfo {title}
  {Localization transition in the anderson model on the bethe lattice:
  spontaneous symmetry breaking and correlation functions},}\ }\href@noop {}
  {\bibfield  {journal} {\bibinfo  {journal} {Nuclear Physics B}\ }\textbf
  {\bibinfo {volume} {366}},\ \bibinfo {pages} {507--532} (\bibinfo {year}
  {1991})}\BibitemShut {NoStop}%
\bibitem [{\citenamefont {Evangelou}(1992)}]{evangelou1992numerical}%
  \BibitemOpen
  \bibfield  {author} {\bibinfo {author} {\bibfnamefont {SN}~\bibnamefont
  {Evangelou}},\ }\bibfield  {title} {\enquote {\bibinfo {title} {A numerical
  study of sparse random matrices},}\ }\href@noop {} {\bibfield  {journal}
  {\bibinfo  {journal} {Journal of statistical physics}\ }\textbf {\bibinfo
  {volume} {69}},\ \bibinfo {pages} {361--383} (\bibinfo {year}
  {1992})}\BibitemShut {NoStop}%
\bibitem [{\citenamefont {Bauer}\ and\ \citenamefont
  {Golinelli}(2001)}]{bauer2001random}%
  \BibitemOpen
  \bibfield  {author} {\bibinfo {author} {\bibfnamefont {Michel}\ \bibnamefont
  {Bauer}}\ and\ \bibinfo {author} {\bibfnamefont {Olivier}\ \bibnamefont
  {Golinelli}},\ }\bibfield  {title} {\enquote {\bibinfo {title} {Random
  incidence matrices: moments of the spectral density},}\ }\href@noop {}
  {\bibfield  {journal} {\bibinfo  {journal} {Journal of Statistical Physics}\
  }\textbf {\bibinfo {volume} {103}},\ \bibinfo {pages} {301--337} (\bibinfo
  {year} {2001})}\BibitemShut {NoStop}%
\bibitem [{\citenamefont {Kühn}(2008)}]{Kuhn2008}%
  \BibitemOpen
  \bibfield  {author} {\bibinfo {author} {\bibfnamefont {Reimer}\ \bibnamefont
  {Kühn}},\ }\bibfield  {title} {\enquote {\bibinfo {title} {Spectra of sparse
  random matrices},}\ }\href {\doibase 10.1088/1751-8113/41/29/295002}
  {\bibfield  {journal} {\bibinfo  {journal} {Journal of Physics A:
  Mathematical and Theoretical}\ }\textbf {\bibinfo {volume} {41}},\ \bibinfo
  {pages} {295002} (\bibinfo {year} {2008})}\BibitemShut {NoStop}%
\bibitem [{\citenamefont {Metz}\ \emph {et~al.}(2010)\citenamefont {Metz},
  \citenamefont {Neri},\ and\ \citenamefont
  {Boll{\'e}}}]{metz2010localization}%
  \BibitemOpen
  \bibfield  {author} {\bibinfo {author} {\bibfnamefont {Fernando~Lucas}\
  \bibnamefont {Metz}}, \bibinfo {author} {\bibfnamefont {Izaak}\ \bibnamefont
  {Neri}}, \ and\ \bibinfo {author} {\bibfnamefont {D{\'e}sir{\'e}}\
  \bibnamefont {Boll{\'e}}},\ }\bibfield  {title} {\enquote {\bibinfo {title}
  {Localization transition in symmetric random matrices},}\ }\href@noop {}
  {\bibfield  {journal} {\bibinfo  {journal} {Physical Review E}\ }\textbf
  {\bibinfo {volume} {82}},\ \bibinfo {pages} {031135} (\bibinfo {year}
  {2010})}\BibitemShut {NoStop}%
\bibitem [{\citenamefont {Kabashima}\ and\ \citenamefont
  {Takahashi}(2012)}]{Kabashima2012}%
  \BibitemOpen
  \bibfield  {author} {\bibinfo {author} {\bibfnamefont {Yoshiyuki}\
  \bibnamefont {Kabashima}}\ and\ \bibinfo {author} {\bibfnamefont {Hisanao}\
  \bibnamefont {Takahashi}},\ }\bibfield  {title} {\enquote {\bibinfo {title}
  {First eigenvalue/eigenvector in sparse random symmetric matrices: influences
  of degree fluctuation},}\ }\href {\doibase 10.1088/1751-8113/45/32/325001}
  {\bibfield  {journal} {\bibinfo  {journal} {Journal of Physics A:
  Mathematical and Theoretical}\ }\textbf {\bibinfo {volume} {45}},\ \bibinfo
  {pages} {325001} (\bibinfo {year} {2012})}\BibitemShut {NoStop}%
\bibitem [{\citenamefont {Slanina}(2012)}]{slanina2012}%
  \BibitemOpen
  \bibfield  {author} {\bibinfo {author} {\bibfnamefont {F.}~\bibnamefont
  {Slanina}},\ }\bibfield  {title} {\enquote {\bibinfo {title} {Localization of
  eigenvectors in random graphs},}\ }\href@noop {} {\bibfield  {journal}
  {\bibinfo  {journal} {Eur. Phys. J. B}\ }\textbf {\bibinfo {volume} {85}},\
  \bibinfo {pages} {361} (\bibinfo {year} {2012})}\BibitemShut {NoStop}%
\bibitem [{\citenamefont {Pastor-Satorras}\ and\ \citenamefont
  {Castellano}(2016)}]{Pastor2016}%
  \BibitemOpen
  \bibfield  {author} {\bibinfo {author} {\bibfnamefont {R.}~\bibnamefont
  {Pastor-Satorras}}\ and\ \bibinfo {author} {\bibfnamefont {C.}~\bibnamefont
  {Castellano}},\ }\bibfield  {title} {\enquote {\bibinfo {title} {Distinct
  types of eigenvector localization in networks},}\ }\href@noop {} {\bibfield
  {journal} {\bibinfo  {journal} {Sci. Rep.}\ }\textbf {\bibinfo {volume}
  {6}},\ \bibinfo {pages} {18847} (\bibinfo {year} {2016})}\BibitemShut
  {NoStop}%
\bibitem [{\citenamefont {Tikhonov}\ \emph {et~al.}(2016)\citenamefont
  {Tikhonov}, \citenamefont {Mirlin},\ and\ \citenamefont
  {Skvortsov}}]{tikhonov2016anderson}%
  \BibitemOpen
  \bibfield  {author} {\bibinfo {author} {\bibfnamefont {KS}~\bibnamefont
  {Tikhonov}}, \bibinfo {author} {\bibfnamefont {AD}~\bibnamefont {Mirlin}}, \
  and\ \bibinfo {author} {\bibfnamefont {MA}~\bibnamefont {Skvortsov}},\
  }\bibfield  {title} {\enquote {\bibinfo {title} {Anderson localization and
  ergodicity on random regular graphs},}\ }\href@noop {} {\bibfield  {journal}
  {\bibinfo  {journal} {Physical Review B}\ }\textbf {\bibinfo {volume} {94}},\
  \bibinfo {pages} {220203} (\bibinfo {year} {2016})}\BibitemShut {NoStop}%
\bibitem [{\citenamefont {Susca}\ \emph {et~al.}(2019)\citenamefont {Susca},
  \citenamefont {Vivo},\ and\ \citenamefont {Kühn}}]{Susca2019}%
  \BibitemOpen
  \bibfield  {author} {\bibinfo {author} {\bibfnamefont {Vito A~R}\
  \bibnamefont {Susca}}, \bibinfo {author} {\bibfnamefont {Pierpaolo}\
  \bibnamefont {Vivo}}, \ and\ \bibinfo {author} {\bibfnamefont {Reimer}\
  \bibnamefont {Kühn}},\ }\bibfield  {title} {\enquote {\bibinfo {title} {Top
  eigenpair statistics for weighted sparse graphs},}\ }\href {\doibase
  10.1088/1751-8121/ab4d63} {\bibfield  {journal} {\bibinfo  {journal} {Journal
  of Physics A: Mathematical and Theoretical}\ }\textbf {\bibinfo {volume}
  {52}},\ \bibinfo {pages} {485002} (\bibinfo {year} {2019})}\BibitemShut
  {NoStop}%
\bibitem [{\citenamefont {Hatano}\ and\ \citenamefont
  {Nelson}(1996)}]{hatano1996localization}%
  \BibitemOpen
  \bibfield  {author} {\bibinfo {author} {\bibfnamefont {Naomichi}\
  \bibnamefont {Hatano}}\ and\ \bibinfo {author} {\bibfnamefont {David~R}\
  \bibnamefont {Nelson}},\ }\bibfield  {title} {\enquote {\bibinfo {title}
  {Localization transitions in non-hermitian quantum mechanics},}\ }\href@noop
  {} {\bibfield  {journal} {\bibinfo  {journal} {Physical review letters}\
  }\textbf {\bibinfo {volume} {77}},\ \bibinfo {pages} {570} (\bibinfo {year}
  {1996})}\BibitemShut {NoStop}%
\bibitem [{\citenamefont {Hatano}\ and\ \citenamefont
  {Nelson}(1997)}]{hatano1997vortex}%
  \BibitemOpen
  \bibfield  {author} {\bibinfo {author} {\bibfnamefont {Naomichi}\
  \bibnamefont {Hatano}}\ and\ \bibinfo {author} {\bibfnamefont {David~R}\
  \bibnamefont {Nelson}},\ }\bibfield  {title} {\enquote {\bibinfo {title}
  {Vortex pinning and non-hermitian quantum mechanics},}\ }\href@noop {}
  {\bibfield  {journal} {\bibinfo  {journal} {Physical Review B}\ }\textbf
  {\bibinfo {volume} {56}},\ \bibinfo {pages} {8651} (\bibinfo {year}
  {1997})}\BibitemShut {NoStop}%
\bibitem [{\citenamefont {Feinberg}\ and\ \citenamefont
  {Zee}(1999)}]{feinberg1999non}%
  \BibitemOpen
  \bibfield  {author} {\bibinfo {author} {\bibfnamefont {Joshua}\ \bibnamefont
  {Feinberg}}\ and\ \bibinfo {author} {\bibfnamefont {A}~\bibnamefont {Zee}},\
  }\bibfield  {title} {\enquote {\bibinfo {title} {Non-hermitian localization
  and delocalization},}\ }\href@noop {} {\bibfield  {journal} {\bibinfo
  {journal} {Physical Review E}\ }\textbf {\bibinfo {volume} {59}},\ \bibinfo
  {pages} {6433} (\bibinfo {year} {1999})}\BibitemShut {NoStop}%
\bibitem [{\citenamefont {Amir}\ \emph {et~al.}(2016)\citenamefont {Amir},
  \citenamefont {Hatano},\ and\ \citenamefont {Nelson}}]{Amir2016}%
  \BibitemOpen
  \bibfield  {author} {\bibinfo {author} {\bibfnamefont {Ariel}\ \bibnamefont
  {Amir}}, \bibinfo {author} {\bibfnamefont {Naomichi}\ \bibnamefont {Hatano}},
  \ and\ \bibinfo {author} {\bibfnamefont {David~R.}\ \bibnamefont {Nelson}},\
  }\bibfield  {title} {\enquote {\bibinfo {title} {Non-hermitian localization
  in biological networks},}\ }\href {\doibase 10.1103/PhysRevE.93.042310}
  {\bibfield  {journal} {\bibinfo  {journal} {Phys. Rev. E}\ }\textbf {\bibinfo
  {volume} {93}},\ \bibinfo {pages} {042310} (\bibinfo {year}
  {2016})}\BibitemShut {NoStop}%
\bibitem [{\citenamefont {Zhang}\ and\ \citenamefont
  {Nelson}(2019)}]{Zhang2019}%
  \BibitemOpen
  \bibfield  {author} {\bibinfo {author} {\bibfnamefont {Grace~H.}\
  \bibnamefont {Zhang}}\ and\ \bibinfo {author} {\bibfnamefont {David~R.}\
  \bibnamefont {Nelson}},\ }\bibfield  {title} {\enquote {\bibinfo {title}
  {Eigenvalue repulsion and eigenvector localization in sparse non-hermitian
  random matrices},}\ }\href {\doibase 10.1103/PhysRevE.100.052315} {\bibfield
  {journal} {\bibinfo  {journal} {Phys. Rev. E}\ }\textbf {\bibinfo {volume}
  {100}},\ \bibinfo {pages} {052315} (\bibinfo {year} {2019})}\BibitemShut
  {NoStop}%
\bibitem [{\citenamefont {Peron}\ \emph {et~al.}(2020)\citenamefont {Peron},
  \citenamefont {de~Resende}, \citenamefont {Rodrigues}, \citenamefont
  {Costa},\ and\ \citenamefont {M\'endez-Berm\'udez}}]{Peron}%
  \BibitemOpen
  \bibfield  {author} {\bibinfo {author} {\bibfnamefont {Thomas}\ \bibnamefont
  {Peron}}, \bibinfo {author} {\bibfnamefont {Bruno Messias~F.}\ \bibnamefont
  {de~Resende}}, \bibinfo {author} {\bibfnamefont {Francisco~A.}\ \bibnamefont
  {Rodrigues}}, \bibinfo {author} {\bibfnamefont {Luciano da~F.}\ \bibnamefont
  {Costa}}, \ and\ \bibinfo {author} {\bibfnamefont {J.~A.}\ \bibnamefont
  {M\'endez-Berm\'udez}},\ }\bibfield  {title} {\enquote {\bibinfo {title}
  {Spacing ratio characterization of the spectra of directed random
  networks},}\ }\href {\doibase 10.1103/PhysRevE.102.062305} {\bibfield
  {journal} {\bibinfo  {journal} {Phys. Rev. E}\ }\textbf {\bibinfo {volume}
  {102}},\ \bibinfo {pages} {062305} (\bibinfo {year} {2020})}\BibitemShut
  {NoStop}%
\bibitem [{\citenamefont {Fosdick}\ \emph {et~al.}(2018)\citenamefont
  {Fosdick}, \citenamefont {Larremore}, \citenamefont {Nishimura},\ and\
  \citenamefont {Ugander}}]{Fosdick}%
  \BibitemOpen
  \bibfield  {author} {\bibinfo {author} {\bibfnamefont {Bailey~K.}\
  \bibnamefont {Fosdick}}, \bibinfo {author} {\bibfnamefont {Daniel~B.}\
  \bibnamefont {Larremore}}, \bibinfo {author} {\bibfnamefont {Joel}\
  \bibnamefont {Nishimura}}, \ and\ \bibinfo {author} {\bibfnamefont {Johan}\
  \bibnamefont {Ugander}},\ }\bibfield  {title} {\enquote {\bibinfo {title}
  {Configuring random graph models with fixed degree sequences},}\ }\href
  {\doibase 10.1137/16M1087175} {\bibfield  {journal} {\bibinfo  {journal}
  {SIAM Review}\ }\textbf {\bibinfo {volume} {60}},\ \bibinfo {pages}
  {315--355} (\bibinfo {year} {2018})},\ \Eprint
  {http://arxiv.org/abs/https://doi.org/10.1137/16M1087175}
  {https://doi.org/10.1137/16M1087175} \BibitemShut {NoStop}%
\bibitem [{\citenamefont {Molloy}\ and\ \citenamefont {Reed}(1995)}]{Molloy95}%
  \BibitemOpen
  \bibfield  {author} {\bibinfo {author} {\bibfnamefont {Michael}\ \bibnamefont
  {Molloy}}\ and\ \bibinfo {author} {\bibfnamefont {Bruce}\ \bibnamefont
  {Reed}},\ }\bibfield  {title} {\enquote {\bibinfo {title} {A critical point
  for random graphs with a given degree sequence},}\ }\href {\doibase
  10.1002/rsa.3240060204} {\bibfield  {journal} {\bibinfo  {journal} {Random
  Structures \& Algorithms}\ }\textbf {\bibinfo {volume} {6}},\ \bibinfo
  {pages} {161--180} (\bibinfo {year} {1995})},\ \Eprint
  {http://arxiv.org/abs/https://onlinelibrary.wiley.com/doi/pdf/10.1002/rsa.3240060204}
  {https://onlinelibrary.wiley.com/doi/pdf/10.1002/rsa.3240060204} \BibitemShut
  {NoStop}%
\bibitem [{\citenamefont {Molloy}\ and\ \citenamefont {Reed}(1998)}]{Molloy98}%
  \BibitemOpen
  \bibfield  {author} {\bibinfo {author} {\bibfnamefont {M}~\bibnamefont
  {Molloy}}\ and\ \bibinfo {author} {\bibfnamefont {B}~\bibnamefont {Reed}},\
  }\bibfield  {title} {\enquote {\bibinfo {title} {The size of the giant
  component of a random graph with a given degree sequence},}\ }\href {\doibase
  10.1017/S0963548398003526} {\bibfield  {journal} {\bibinfo  {journal}
  {Combinatorics, Probability and Computing}\ }\textbf {\bibinfo {volume}
  {7}},\ \bibinfo {pages} {295–305} (\bibinfo {year} {1998})}\BibitemShut
  {NoStop}%
\bibitem [{\citenamefont {Bollob{\'a}s}\ and\ \citenamefont
  {B{\'e}la}(2001)}]{bollobas2001random}%
  \BibitemOpen
  \bibfield  {author} {\bibinfo {author} {\bibfnamefont {B{\'e}la}\
  \bibnamefont {Bollob{\'a}s}}\ and\ \bibinfo {author} {\bibfnamefont
  {Bollob{\'a}s}\ \bibnamefont {B{\'e}la}},\ }\href@noop {} {\emph {\bibinfo
  {title} {Random graphs}}},\ \bibinfo {number} {73}\ (\bibinfo  {publisher}
  {Cambridge university press},\ \bibinfo {year} {2001})\BibitemShut {NoStop}%
\bibitem [{\citenamefont {Newman}\ \emph {et~al.}(2001)\citenamefont {Newman},
  \citenamefont {Strogatz},\ and\ \citenamefont {Watts}}]{Newman2001}%
  \BibitemOpen
  \bibfield  {author} {\bibinfo {author} {\bibfnamefont {M.~E.~J.}\
  \bibnamefont {Newman}}, \bibinfo {author} {\bibfnamefont {S.~H.}\
  \bibnamefont {Strogatz}}, \ and\ \bibinfo {author} {\bibfnamefont {D.~J.}\
  \bibnamefont {Watts}},\ }\bibfield  {title} {\enquote {\bibinfo {title}
  {Random graphs with arbitrary degree distributions and their applications},}\
  }\href {\doibase 10.1103/PhysRevE.64.026118} {\bibfield  {journal} {\bibinfo
  {journal} {Phys. Rev. E}\ }\textbf {\bibinfo {volume} {64}},\ \bibinfo
  {pages} {026118} (\bibinfo {year} {2001})}\BibitemShut {NoStop}%
\bibitem [{\citenamefont {Newman}(2010)}]{Newman10}%
  \BibitemOpen
  \bibfield  {author} {\bibinfo {author} {\bibfnamefont {M.}~\bibnamefont
  {Newman}},\ }\href {https://books.google.com.br/books?id=q7HVtpYVfC0C} {\emph
  {\bibinfo {title} {Networks: An Introduction}}}\ (\bibinfo  {publisher} {OUP
  Oxford},\ \bibinfo {year} {2010})\BibitemShut {NoStop}%
\bibitem [{\citenamefont {Dorogovtsev}\ and\ \citenamefont
  {Mendes}(2013)}]{dorogovtsev2013evolution}%
  \BibitemOpen
  \bibfield  {author} {\bibinfo {author} {\bibfnamefont {Sergei~N}\
  \bibnamefont {Dorogovtsev}}\ and\ \bibinfo {author} {\bibfnamefont
  {Jos{\'e}~FF}\ \bibnamefont {Mendes}},\ }\href@noop {} {\emph {\bibinfo
  {title} {Evolution of networks: From biological nets to the Internet and
  WWW}}}\ (\bibinfo  {publisher} {OUP Oxford},\ \bibinfo {year}
  {2013})\BibitemShut {NoStop}%
\bibitem [{\citenamefont {Arbib}(2003)}]{amari2003handbook}%
  \BibitemOpen
  \bibinfo {editor} {\bibfnamefont {Michael~A}\ \bibnamefont {Arbib}},\ ed.,\
  \href@noop {} {\emph {\bibinfo {title} {The handbook of brain theory and
  neural networks}}}\ (\bibinfo  {publisher} {MIT press},\ \bibinfo {year}
  {2003})\BibitemShut {NoStop}%
\bibitem [{\citenamefont {Rogers}\ and\ \citenamefont
  {Castillo}(2009)}]{Rogers2009}%
  \BibitemOpen
  \bibfield  {author} {\bibinfo {author} {\bibfnamefont {Tim}\ \bibnamefont
  {Rogers}}\ and\ \bibinfo {author} {\bibfnamefont {Isaac~P\'erez}\
  \bibnamefont {Castillo}},\ }\bibfield  {title} {\enquote {\bibinfo {title}
  {Cavity approach to the spectral density of non-hermitian sparse matrices},}\
  }\href {\doibase 10.1103/PhysRevE.79.012101} {\bibfield  {journal} {\bibinfo
  {journal} {Phys. Rev. E}\ }\textbf {\bibinfo {volume} {79}},\ \bibinfo
  {pages} {012101} (\bibinfo {year} {2009})}\BibitemShut {NoStop}%
\bibitem [{\citenamefont {Neri}\ and\ \citenamefont {Metz}(2016)}]{Neri2016}%
  \BibitemOpen
  \bibfield  {author} {\bibinfo {author} {\bibfnamefont {Izaak}\ \bibnamefont
  {Neri}}\ and\ \bibinfo {author} {\bibfnamefont {Fernando~Lucas}\ \bibnamefont
  {Metz}},\ }\bibfield  {title} {\enquote {\bibinfo {title} {Eigenvalue
  outliers of non-hermitian random matrices with a local tree structure},}\
  }\href {\doibase 10.1103/PhysRevLett.117.224101} {\bibfield  {journal}
  {\bibinfo  {journal} {Phys. Rev. Lett.}\ }\textbf {\bibinfo {volume} {117}},\
  \bibinfo {pages} {224101} (\bibinfo {year} {2016})}\BibitemShut {NoStop}%
\bibitem [{\citenamefont {Metz}\ \emph {et~al.}(2019)\citenamefont {Metz},
  \citenamefont {Neri},\ and\ \citenamefont {Rogers}}]{Metz2019}%
  \BibitemOpen
  \bibfield  {author} {\bibinfo {author} {\bibfnamefont {Fernando~Lucas}\
  \bibnamefont {Metz}}, \bibinfo {author} {\bibfnamefont {Izaak}\ \bibnamefont
  {Neri}}, \ and\ \bibinfo {author} {\bibfnamefont {Tim}\ \bibnamefont
  {Rogers}},\ }\bibfield  {title} {\enquote {\bibinfo {title} {Spectral theory
  of sparse non-hermitian random matrices},}\ }\href {\doibase
  10.1088/1751-8121/ab1ce0} {\bibfield  {journal} {\bibinfo  {journal} {Journal
  of Physics A: Mathematical and Theoretical}\ }\textbf {\bibinfo {volume}
  {52}},\ \bibinfo {pages} {434003} (\bibinfo {year} {2019})}\BibitemShut
  {NoStop}%
\bibitem [{\citenamefont {Neri}\ and\ \citenamefont {Metz}(2020)}]{Neri2019}%
  \BibitemOpen
  \bibfield  {author} {\bibinfo {author} {\bibfnamefont {Izaak}\ \bibnamefont
  {Neri}}\ and\ \bibinfo {author} {\bibfnamefont {Fernando~Lucas}\ \bibnamefont
  {Metz}},\ }\bibfield  {title} {\enquote {\bibinfo {title} {Linear stability
  analysis of large dynamical systems on random directed graphs},}\ }\href
  {\doibase 10.1103/PhysRevResearch.2.033313} {\bibfield  {journal} {\bibinfo
  {journal} {Phys. Rev. Research}\ }\textbf {\bibinfo {volume} {2}},\ \bibinfo
  {pages} {033313} (\bibinfo {year} {2020})}\BibitemShut {NoStop}%
\bibitem [{\citenamefont {Dorogovtsev}\ \emph {et~al.}(2001)\citenamefont
  {Dorogovtsev}, \citenamefont {Mendes},\ and\ \citenamefont
  {Samukhin}}]{dorogovtsev2001giant}%
  \BibitemOpen
  \bibfield  {author} {\bibinfo {author} {\bibfnamefont {S.~N.}\ \bibnamefont
  {Dorogovtsev}}, \bibinfo {author} {\bibfnamefont {J.~F.~F.}\ \bibnamefont
  {Mendes}}, \ and\ \bibinfo {author} {\bibfnamefont {A.~N.}\ \bibnamefont
  {Samukhin}},\ }\bibfield  {title} {\enquote {\bibinfo {title} {Giant strongly
  connected component of directed networks},}\ }\href {\doibase
  10.1103/PhysRevE.64.025101} {\bibfield  {journal} {\bibinfo  {journal} {Phys.
  Rev. E}\ }\textbf {\bibinfo {volume} {64}},\ \bibinfo {pages} {025101}
  (\bibinfo {year} {2001})}\BibitemShut {NoStop}%
\bibitem [{\citenamefont {Fyodorov}\ and\ \citenamefont
  {Mirlin}(1994)}]{Mirlin1994}%
  \BibitemOpen
  \bibfield  {author} {\bibinfo {author} {\bibfnamefont {Y~V}\ \bibnamefont
  {Fyodorov}}\ and\ \bibinfo {author} {\bibfnamefont {A~D}\ \bibnamefont
  {Mirlin}},\ }\bibfield  {title} {\enquote {\bibinfo {title} {Statistical
  properties of eigenfunctions of random quasi 1d one-particle hamiltonians},}\
  }\href@noop {} {\bibfield  {journal} {\bibinfo  {journal} {Int. J. Mod. Phys.
  B}\ }\textbf {\bibinfo {volume} {8}},\ \bibinfo {pages} {3795--3842}
  (\bibinfo {year} {1994})}\BibitemShut {NoStop}%
\bibitem [{\citenamefont {Efetov}(1999)}]{efetov1999supersymmetry}%
  \BibitemOpen
  \bibfield  {author} {\bibinfo {author} {\bibfnamefont {Konstantin}\
  \bibnamefont {Efetov}},\ }\href@noop {} {\emph {\bibinfo {title}
  {Supersymmetry in disorder and chaos}}}\ (\bibinfo  {publisher} {Cambridge
  University Press},\ \bibinfo {year} {1999})\BibitemShut {NoStop}%
\bibitem [{\citenamefont {Metz}\ and\ \citenamefont {Neri}(2020)}]{SM}%
  \BibitemOpen
  \bibfield  {author} {\bibinfo {author} {\bibfnamefont {F.~L.}\ \bibnamefont
  {Metz}}\ and\ \bibinfo {author} {\bibfnamefont {I.}~\bibnamefont {Neri}},\
  }\bibfield  {title} {\enquote {\bibinfo {title} {see the supplemental
  material},}\ }\href@noop {} {\bibfield  {journal} {\bibinfo  {journal} {See
  Supplemental Material}\ } (\bibinfo {year} {2020})}\BibitemShut {NoStop}%
\bibitem [{\citenamefont {Porter}\ and\ \citenamefont
  {Thomas}(1956)}]{Porter1956}%
  \BibitemOpen
  \bibfield  {author} {\bibinfo {author} {\bibfnamefont {C.~E.}\ \bibnamefont
  {Porter}}\ and\ \bibinfo {author} {\bibfnamefont {R.~G.}\ \bibnamefont
  {Thomas}},\ }\bibfield  {title} {\enquote {\bibinfo {title} {Fluctuations of
  nuclear reaction widths},}\ }\href {\doibase 10.1103/PhysRev.104.483}
  {\bibfield  {journal} {\bibinfo  {journal} {Phys. Rev.}\ }\textbf {\bibinfo
  {volume} {104}},\ \bibinfo {pages} {483--491} (\bibinfo {year}
  {1956})}\BibitemShut {NoStop}%
\bibitem [{\citenamefont {Mirlin}(2000)}]{Mirlin2000}%
  \BibitemOpen
  \bibfield  {author} {\bibinfo {author} {\bibfnamefont {Alexander~D.}\
  \bibnamefont {Mirlin}},\ }\bibfield  {title} {\enquote {\bibinfo {title}
  {Statistics of energy levels and eigenfunctions in disordered systems},}\
  }\href {\doibase https://doi.org/10.1016/S0370-1573(99)00091-5} {\bibfield
  {journal} {\bibinfo  {journal} {Physics Reports}\ }\textbf {\bibinfo {volume}
  {326}},\ \bibinfo {pages} {259 -- 382} (\bibinfo {year} {2000})}\BibitemShut
  {NoStop}%
\bibitem [{\citenamefont {Gradshteyn}\ and\ \citenamefont
  {Ryzhik}(2014)}]{grad2007}%
  \BibitemOpen
  \bibfield  {author} {\bibinfo {author} {\bibfnamefont {I.S.}\ \bibnamefont
  {Gradshteyn}}\ and\ \bibinfo {author} {\bibfnamefont {I.M.}\ \bibnamefont
  {Ryzhik}},\ }\href {https://books.google.com.br/books?id=F7jiBQAAQBAJ} {\emph
  {\bibinfo {title} {Table of Integrals, Series, and Products}}}\ (\bibinfo
  {publisher} {Elsevier Science},\ \bibinfo {year} {2014})\BibitemShut
  {NoStop}%
\bibitem [{\citenamefont {Krivelevich}\ and\ \citenamefont
  {Sudakov}(2003)}]{krivelevich2003largest}%
  \BibitemOpen
  \bibfield  {author} {\bibinfo {author} {\bibfnamefont {Michael}\ \bibnamefont
  {Krivelevich}}\ and\ \bibinfo {author} {\bibfnamefont {Benny}\ \bibnamefont
  {Sudakov}},\ }\bibfield  {title} {\enquote {\bibinfo {title} {The largest
  eigenvalue of sparse random graphs},}\ }\href {\doibase
  10.1017/S0963548302005424} {\bibfield  {journal} {\bibinfo  {journal}
  {Combinatorics, Probability and Computing}\ }\textbf {\bibinfo {volume}
  {12}},\ \bibinfo {pages} {61–72} (\bibinfo {year} {2003})}\BibitemShut
  {NoStop}%
\bibitem [{\citenamefont {Nadakuditi}\ and\ \citenamefont
  {Newman}(2013)}]{Nada2013}%
  \BibitemOpen
  \bibfield  {author} {\bibinfo {author} {\bibfnamefont {Raj~Rao}\ \bibnamefont
  {Nadakuditi}}\ and\ \bibinfo {author} {\bibfnamefont {M.~E.~J.}\ \bibnamefont
  {Newman}},\ }\bibfield  {title} {\enquote {\bibinfo {title} {Spectra of
  random graphs with arbitrary expected degrees},}\ }\href {\doibase
  10.1103/PhysRevE.87.012803} {\bibfield  {journal} {\bibinfo  {journal} {Phys.
  Rev. E}\ }\textbf {\bibinfo {volume} {87}},\ \bibinfo {pages} {012803}
  (\bibinfo {year} {2013})}\BibitemShut {NoStop}%
\bibitem [{\citenamefont {Metz}\ and\ \citenamefont {Silva}(2020)}]{MetzPRR}%
  \BibitemOpen
  \bibfield  {author} {\bibinfo {author} {\bibfnamefont {Fernando~L.}\
  \bibnamefont {Metz}}\ and\ \bibinfo {author} {\bibfnamefont {Jeferson~D.}\
  \bibnamefont {Silva}},\ }\bibfield  {title} {\enquote {\bibinfo {title}
  {Spectral density of dense random networks and the breakdown of the wigner
  semicircle law},}\ }\href {\doibase 10.1103/PhysRevResearch.2.043116}
  {\bibfield  {journal} {\bibinfo  {journal} {Phys. Rev. Research}\ }\textbf
  {\bibinfo {volume} {2}},\ \bibinfo {pages} {043116} (\bibinfo {year}
  {2020})}\BibitemShut {NoStop}%
\bibitem [{\citenamefont {Grilli}\ \emph {et~al.}(2016)\citenamefont {Grilli},
  \citenamefont {Rogers},\ and\ \citenamefont {Allesina}}]{Grilli2016}%
  \BibitemOpen
  \bibfield  {author} {\bibinfo {author} {\bibfnamefont {Jacopo}\ \bibnamefont
  {Grilli}}, \bibinfo {author} {\bibfnamefont {Tim}\ \bibnamefont {Rogers}}, \
  and\ \bibinfo {author} {\bibfnamefont {Stefano}\ \bibnamefont {Allesina}},\
  }\bibfield  {title} {\enquote {\bibinfo {title} {Modularity and stability in
  ecological communities},}\ }\href@noop {} {\bibfield  {journal} {\bibinfo
  {journal} {Nat. Commun.}\ }\textbf {\bibinfo {volume} {7}},\ \bibinfo {pages}
  {12031} (\bibinfo {year} {2016})}\BibitemShut {NoStop}%
\bibitem [{\citenamefont {Dorogovtsev}\ \emph {et~al.}(2003)\citenamefont
  {Dorogovtsev}, \citenamefont {Goltsev}, \citenamefont {Mendes},\ and\
  \citenamefont {Samukhin}}]{dorogovtsev2003spectra}%
  \BibitemOpen
  \bibfield  {author} {\bibinfo {author} {\bibfnamefont {Sergey~N}\
  \bibnamefont {Dorogovtsev}}, \bibinfo {author} {\bibfnamefont {Alexander~V}\
  \bibnamefont {Goltsev}}, \bibinfo {author} {\bibfnamefont {Jos{\'e}~FF}\
  \bibnamefont {Mendes}}, \ and\ \bibinfo {author} {\bibfnamefont
  {Alexander~N}\ \bibnamefont {Samukhin}},\ }\bibfield  {title} {\enquote
  {\bibinfo {title} {Spectra of complex networks},}\ }\href@noop {} {\bibfield
  {journal} {\bibinfo  {journal} {Physical Review E}\ }\textbf {\bibinfo
  {volume} {68}},\ \bibinfo {pages} {046109} (\bibinfo {year}
  {2003})}\BibitemShut {NoStop}%
\bibitem [{\citenamefont {St{\"a}ring}\ \emph {et~al.}(2003)\citenamefont
  {St{\"a}ring}, \citenamefont {Mehlig}, \citenamefont {Fyodorov},\ and\
  \citenamefont {Luck}}]{staring2003random}%
  \BibitemOpen
  \bibfield  {author} {\bibinfo {author} {\bibfnamefont {J}~\bibnamefont
  {St{\"a}ring}}, \bibinfo {author} {\bibfnamefont {B}~\bibnamefont {Mehlig}},
  \bibinfo {author} {\bibfnamefont {Yan~V}\ \bibnamefont {Fyodorov}}, \ and\
  \bibinfo {author} {\bibfnamefont {JM}~\bibnamefont {Luck}},\ }\bibfield
  {title} {\enquote {\bibinfo {title} {Random symmetric matrices with a
  constraint: The spectral density of random impedance networks},}\ }\href@noop
  {} {\bibfield  {journal} {\bibinfo  {journal} {Physical Review E}\ }\textbf
  {\bibinfo {volume} {67}},\ \bibinfo {pages} {047101} (\bibinfo {year}
  {2003})}\BibitemShut {NoStop}%
\bibitem [{\citenamefont {Samukhin}\ \emph {et~al.}(2008)\citenamefont
  {Samukhin}, \citenamefont {Dorogovtsev},\ and\ \citenamefont
  {Mendes}}]{samukhin2008laplacian}%
  \BibitemOpen
  \bibfield  {author} {\bibinfo {author} {\bibfnamefont {AN}~\bibnamefont
  {Samukhin}}, \bibinfo {author} {\bibfnamefont {SN}~\bibnamefont
  {Dorogovtsev}}, \ and\ \bibinfo {author} {\bibfnamefont {JFF}\ \bibnamefont
  {Mendes}},\ }\bibfield  {title} {\enquote {\bibinfo {title} {Laplacian
  spectra of, and random walks on, complex networks: Are scale-free
  architectures really important?}}\ }\href@noop {} {\bibfield  {journal}
  {\bibinfo  {journal} {Physical Review E}\ }\textbf {\bibinfo {volume} {77}},\
  \bibinfo {pages} {036115} (\bibinfo {year} {2008})}\BibitemShut {NoStop}%
\bibitem [{\citenamefont {K{\"u}hn}(2015)}]{kuhn2015spectra}%
  \BibitemOpen
  \bibfield  {author} {\bibinfo {author} {\bibfnamefont {Reimer}\ \bibnamefont
  {K{\"u}hn}},\ }\bibfield  {title} {\enquote {\bibinfo {title} {Spectra of
  random stochastic matrices and relaxation in complex systems},}\ }\href@noop
  {} {\bibfield  {journal} {\bibinfo  {journal} {EPL (Europhysics Letters)}\
  }\textbf {\bibinfo {volume} {109}},\ \bibinfo {pages} {60003} (\bibinfo
  {year} {2015})}\BibitemShut {NoStop}%
\bibitem [{\citenamefont {van Mieghem}(2012)}]{vanm2012}%
  \BibitemOpen
  \bibfield  {author} {\bibinfo {author} {\bibfnamefont {P.}~\bibnamefont {van
  Mieghem}},\ }\href {https://books.google.com.br/books?id=hwJFMAEACAAJ} {\emph
  {\bibinfo {title} {Graph Spectra for Complex Networks}}}\ (\bibinfo
  {publisher} {Cambridge University Press},\ \bibinfo {year}
  {2012})\BibitemShut {NoStop}%
\bibitem [{\citenamefont {Borel}(1942)}]{Borel1942}%
  \BibitemOpen
  \bibfield  {author} {\bibinfo {author} {\bibfnamefont {E.}~\bibnamefont
  {Borel}},\ }\href@noop {} {\bibfield  {journal} {\bibinfo  {journal} {C. R.
  Acad. Sci.}\ }\textbf {\bibinfo {volume} {214}},\ \bibinfo {pages} {452}
  (\bibinfo {year} {1942})}\BibitemShut {NoStop}%
\bibitem [{\citenamefont {Tanner}(1961)}]{Tanner1961}%
  \BibitemOpen
  \bibfield  {author} {\bibinfo {author} {\bibfnamefont {J.~C.}\ \bibnamefont
  {Tanner}},\ }\bibfield  {title} {\enquote {\bibinfo {title} {{A derivation of
  the Borel distribution}},}\ }\href {\doibase 10.1093/biomet/48.1-2.222}
  {\bibfield  {journal} {\bibinfo  {journal} {Biometrika}\ }\textbf {\bibinfo
  {volume} {48}},\ \bibinfo {pages} {222--224} (\bibinfo {year} {1961})},\
  \Eprint
  {http://arxiv.org/abs/https://academic.oup.com/biomet/article-pdf/48/1-2/222/607732/48-1-2-222.pdf}
  {https://academic.oup.com/biomet/article-pdf/48/1-2/222/607732/48-1-2-222.pdf}
  \BibitemShut {NoStop}%
\end{thebibliography}%

\clearpage
\setboolean{@twoside}{false}


\section*{Supplementary material}

# Supplemental material for "Localization and universality of eigenvectors in directed random graphs"

Fernando Lucas Metz

*Physics Institute, Federal University of Rio Grande do Sul, 91501-970 Porto Alegre, Brazil and*

*London Mathematical Laboratory, 18 Margravine Gardens, London W6 8RH, United Kingdom*

Izaak Neri

*Department of Mathematics, Kings College London, Strand, London, WC2R 2LS, UK*

(Dated: December 7, 2020)

## S1. INTRODUCTION

The supplements are organized into four sections. In Sec. S2, we discuss the degree distributions we consider in the paper: regular, Poisson, exponential and Borel distributions. In Sec. S3, we derive Eqs. (9) and (10) in the main text for the inverse participation ratios $\mathcal{I}(\lambda_{\rm b})$ and $\mathcal{I}(\lambda_{\rm isol})$, respectively. In Sec. S4, we derive the analytic results for the distribution of the eigenvector components in the high connectivity limit. Lastly, in Sec. S5, we compare theoretical results for the inverse participation ratio of infinitely large matrices with numerical results for matrices of finite size.

## S2. DIRECTED RANDOM GRAPHS WITH A PRESCRIBED DEGREE DISTRIBUTION

In the present paper, we consider random graphs with a prescribed degree distribution

$$p_{K^{\rm in},K^{\rm out}}(k,\ell) = p_{K^{\rm in}}(k) p_{K^{\rm out}}(\ell) \tag{S1}$$

of indegrees $K^{\rm in}$ and outdegrees $K^{\rm out}$, see Refs. [1–5]. Random graph models, in which the degree sequences are specified at the outset, are also called configuration models [1, 6]. We derive results for the spectral properties of simple graphs where self-edges and multi-edges are absent. A simple, weighted, and directed random graph instance, with degrees from a prescribed degree distribution $p_{K^{\rm in},K^{\rm out}}$, can be generated by adapting the standard

stub matching procedure [7] to directed graphs. First, we draw a sequence of indegrees $(K_1^{\rm in}, K_2^{\rm in}, \ldots, K_n^{\rm in})$ and outdegrees $(K_1^{\rm out}, K_2^{\rm out}, \ldots, K_n^{\rm out})$ from the distribution $p_{K^{\rm in}, K^{\rm out}}$, Eq. (S1), conditioned by the constraint $\sum_{j=1}^n K_j^{\rm in} = \sum_{j=1}^n K_j^{\rm out}$. Second, we assign a total number of $K_i^{\rm out}$ outward stubs and $K_i^{\rm in}$ inward stubs to each node $i$ ($i = 1, \ldots, N$). A pair of stubs, containing an outward stub and an inward stub, is selected at random with uniform probability and then connected to create a directed edge between a pair of nodes. The corresponding entry of the adjacency matrix $\boldsymbol{A}$ is set to a value drawn from the distribution of weights $p_J$. This stub matching process is repeated until there are no remainder stubs and a simple graph is returned, i.e., graph instances containing self-edges or multiedges are rejected. This procedure generates directed random graphs that are uniformly sampled from the space of simple graphs. We point out that, although we only consider simple graphs, the results for the spectral properties of $\boldsymbol{A}$ presented in this paper should also apply to the configuration model of directed graphs with multiedges and self-edges. The main reason is that the fraction of multiedges and self-edges vanishes when $N \to \infty$ [4], provided the ratio $c/N$ goes to zero as $N \to \infty$. The latter condition is satisfied if $c$ scales very slowly with $N$.

Because nodes are connected in a random fashion, the local neighbourhood of a randomly selected node is with probability one locally tree-like and oriented in the limit $n \to \infty$. The oriented property means that all edges are unidirectional. The relevance of the oriented property and the local tree-like structure of the graph to the derivation of Eq. (7) in the main text has been discussed in previous works [8, 9].

Below we present some properties of four examples of degree distributions, namely, regular, Poisson, exponential, and Borel distributions. The properties of these distributions are important to reproduce and understand some of the results in the main text. We will only specify the outdegree distribution $p_{K^{\rm out}}(k)$, since $p_{K^{\rm out}} = (k) p_{K^{\rm in}}(k)$ in each case.

### A. Regular graphs

In the $c$-regular ensemble, the indegree and outdegree of each node is equal to a constant $c$ [10], such that

$$p_{K^{\rm out}}(k) = p_{K^{\rm in}}(k) = \delta_{k,c}, \tag{S2}$$

where $\delta$ is the Kronecker delta function. The moments of $p_{K^{\rm out}}(k)$ read

$$\langle (K^{\rm out})^n \rangle = c^n, \quad n \geq 1, \tag{S3}$$

while the variance of the outdegree distribution is given by

$$\text{var}[K^{\text{out}}] = 0. \tag{S4}$$

### B. Poisson graphs

For this ensemble, the degree distribution is given by

$$p_{K^{\text{out}}}(k) = \frac{e^{-c}c^k}{k!}, \tag{S5}$$

where $c > 0$ is a real parameter. Analytic expressions for the moments follow from the generating function

$$g(x) = \sum_{k=0}^{\infty} p_{K^{\text{out}}}(k)e^{xk} = e^{c(e^x-1)} \tag{S6}$$

through the derivatives

$$\langle (K^{\text{out}})^n \rangle = \frac{d^n g}{dx^n}\bigg|_{x=0}. \tag{S7}$$

In the main text of the paper, we need the first four moments

$$\begin{aligned}
\langle K^{\text{out}} \rangle &= c, \\
\langle (K^{\text{out}})^2 \rangle &= c(c+1), \\
\langle (K^{\text{out}})^3 \rangle &= c(c+1) + c^2(c+2), \\
\langle (K^{\text{out}})^4 \rangle &= c(c+1) + 3c^2(c+2) + c^3(c+3).
\end{aligned} \tag{S8}$$

The variance of the outdegree distribution is

$$\text{var}[K^{\text{out}}] = c. \tag{S9}$$

### C. Exponential graphs

For exponential graphs, the degree distribution is

$$p_{K^{\text{out}}}(k) = \frac{1}{c+1}\left(\frac{c}{c+1}\right)^k, \tag{S10}$$

with $c > 0$. By making the change of variables $x \equiv \ln\left(\frac{c}{c+1}\right)$, the $n$-moment can be computed from the equation

$$\langle (K^{\text{out}})^n \rangle = \frac{1}{c+1}\sum_{k=0}^{\infty} k^n e^{xk} = \frac{1}{c+1}\left[\frac{d^n}{dx^n}\frac{1}{1-e^x}\right]\bigg|_{x=\ln\left(\frac{c}{c+1}\right)}. \tag{S11}$$

The analytic expressions for the first four moments read

$$\langle K^{\text{out}} \rangle = c,$$
$$\langle (K^{\text{out}})^2 \rangle = 2c^2 + c,$$
$$\langle (K^{\text{out}})^3 \rangle = 6c^3 + 6c^2 + c,$$
$$\langle (K^{\text{out}})^4 \rangle = 24c^4 + 36c^3 + 14c^2 + c,$$

and the variance is given by

$$\text{var}[K^{\text{out}}] = c^2 + c. \tag{S12}$$

### D. Borel graphs

In these graphs $p_{K^{\text{out}}}(k)$ follows a Borel distribution. The Borel distribution is defined as [11, 12]

$$p_{K^{\text{out}}}(k) = \frac{e^{-\mu k} (\mu k)^{k-1}}{k!}, \tag{S13}$$

where $k$ is a positive integer and $\mu \in [0, 1]$ is a control parameter. As far as we are aware, the Borel degree distribution has not been considered in the context of random networks, thus we discuss in more detail the calculation of the moments in this case.

The moments of the Borel distribution can be evaluated in a recursive way. Here we lay out this recursive approach explicitly for the first two moments, while the third and the fourth moments follow through a similar calculation which we do not present explicitly. We use the change of variables

$$\nu \equiv \mu e^{-\mu} \tag{S14}$$

in the normalization condition $\sum_{k=1}^{\infty} p_{K^{\text{out}}}(k) = 1$ to obtain the useful identity

$$\sum_{k=1}^{\infty} \frac{k^{k-1} \nu^k}{k!} = \mu. \tag{S15}$$

We first address how $\mu$ is related to the mean outdegree

$$c = \langle K^{\text{out}} \rangle. \tag{S16}$$

We take the derivative of Eq. (S15) with respect to $\nu$ and obtain

$$c = \frac{\nu}{\mu} \frac{d\mu}{d\nu}. \tag{S17}$$

The derivative $\frac{d\mu}{d\nu}$ follows from Eq. (S14), namely,

$$\frac{d\mu}{d\nu} = \frac{e^\mu}{1-\mu}. \tag{S18}$$

Substituting Eqs. (S14) and (S18) in Eq. (S17) leads to

$$c = \frac{1}{1-\mu}. \tag{S19}$$

Hence, analogously to the Poisson and exponential outdegree distributions, $p_{K^{\text{out}}}(k)$ is parametrized solely by the mean outdegree $c$ through the relation (S19). The limit $\mu \to 1$ corresponds to the high connectivity limit $c \to \infty$, while $\mu \to 0$ yields $c \to 1$.

The second moment of the outdegree distribution is given by

$$\langle (K^{\text{out}})^2 \rangle = \sum_{k=1}^{\infty} k^2 p_{K^{\text{out}}}(k) = \sum_{k=1}^{\infty} \frac{k^{k+1}\nu^k}{\mu k!}. \tag{S20}$$

By taking a second order derivative of Eq. (S15) with respect to $\nu$ and using the above equation, we obtain

$$\langle (K^{\text{out}})^2 \rangle = \frac{\nu^2}{\mu} \frac{d^2\mu}{d\nu^2} + c. \tag{S21}$$

The expression for $\frac{d^2\mu}{d\nu^2}$ is obtained from taking the derivative of Eq. (S18) with respect to $\nu$, namely,

$$\frac{d^2\mu}{d\nu^2} = \frac{e^{2\mu}(2-\mu)}{(1-\mu)^3}. \tag{S22}$$

We get the analytic expression for $\langle (K^{\text{out}})^2 \rangle$ by plugging the above equation back in Eq. (S21), yielding

$$\langle (K^{\text{out}})^2 \rangle = \frac{1}{(1-\mu)^3}. \tag{S23}$$

The third and fourth moments of $p_{K^{\text{out}}}(k)$ are computed following the same recursive approach. The final analytic expressions are given by

$$\langle (K^{\text{out}})^3 \rangle = \frac{1+2\mu}{(1-\mu)^5} \tag{S24}$$

and

$$\langle (K^{\text{out}})^4 \rangle = \frac{-12\mu^3 + 24\mu^2 + 8\mu - 5}{(1-\mu)^7}. \tag{S25}$$

The variance of the Borel outdegree distribution reads

$$\text{var}[K^{\text{out}}] = c^3 - c^2. \tag{S26}$$

## E. Comparing the degree distributions

The four outdegree distributions introduced above have finite moments and are parametrized by a single parameter, namely the mean outdegree $c$. For a fixed $c$, we can compare four outdegree distributions with the same mean $c$ but different variances. Thus, we use these four degree distributions to illustrate how degree fluctuations affect the localization and the universality properties of the eigenvectors of directed random graphs.

Figure S1 shows the Poisson, the exponential, and the Borel outdegree distributions for $c = 5$. The Borel distribution $p_{K^{\rm out}}(k)$ decays slower for $k \gg 1$ in comparison to the other distributions, i.e., the probability of having nodes with large outdegree $k$ is higher for the Borel distribution than for regular, Poisson, and exponential distributions.

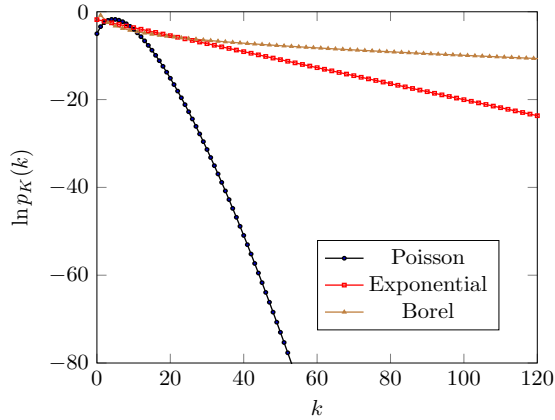

FIG. S1: Poisson, exponential, and Borel outdegree distributions $p_{K^{\rm out}}(k)$ for $c = 5$ (see Eqs. (S5), (S10), and (S13)). The $y$-axis displays the logarithm of $p_{K^{\rm out}}(k)$ to highlight the difference in the tails of the distributions.

The strength of the degree fluctuations of the four ensembles is also captured by the scaling of the variance $\sigma^2_{K^{\rm out}}$ with $c \gg 1$. In all four cases, the variance of $K^{\rm out}$ scales as

$$\text{var}[K^{\rm out}] = Bc^\alpha, \tag{S27}$$

for $c \gg 1$. In particular, for regular graphs $\alpha \to -\infty$, for Poisson graphs $\alpha = 1$, for exponential graphs $\alpha = 2$, and for Borelian graphs $\alpha = 3$. According to Eqs. (15) and (16) in the main text, the different universal behaviors of the right eigenvector distributions for $c \to \infty$ can be classified in terms of the asymptotic behaviour of the relative variance

var$[K^{\rm out}]/c^2$, which is essentially controlled by the exponent $\alpha$. The parameter var$[K^{\rm out}]/c^2$ vanishes for regular random graphs and for the Poisson distribution (both models have $\alpha < 2$), var$[K^{\rm out}]/c^2$ diverges for the Borel distribution ($\alpha > 2$), and var$[K^{\rm out}]/c^2$ remains finite for the exponential distribution ($\alpha = 2$). In the marginal case of $\alpha = 2$, the high connectivity limit of the eigenvector distribution is determined by the prefactor $B$ ($B = 1$ for the exponential distribution). Each of the outdegree distributions discussed here provides an example of a network ensemble within a certain universality class (see also table I in the main text).

## S3. CALCULATION OF THE INVERSE PARTICIPATION RATIO

We derive Eqs. (9) and (10) in the main text for the inverse participation ratio (IPR) of the right eigenvectors associated to the outlier $\lambda_{\rm isol}$ and to an eigenvalue $\lambda_{\rm b}$ at the boundary of the spectral distribution $\rho_{\mathbf{A}}(\lambda)$. We restrict ourselves to right eigenvectors since the results for left eigenvectors are obtained through the substitution "out $\to$ in". The IPR $\mathcal{I}(\lambda)$ of a right eigenvector $\vec{R}(\lambda)$ is determined from (see also Eq. (8) in the main text)

$$\mathcal{I}(\lambda) = \frac{\langle |R(\lambda)|^4 \rangle}{\langle |R(\lambda)|^2 \rangle^2}. \tag{S28}$$

Below we show how to compute the moments $\langle |R(\lambda)|^2 \rangle$ and $\langle |R(\lambda)|^4 \rangle$ of the distribution $p_R$ of the right eigenvector components.

### A. Moments of $p_R$

As shown in Refs. [8, 13], the limit $n \to \infty$ of the distribution $p_R(r|\lambda)$ of the right eigenvector components associated to an eigenvalue $\lambda = \lambda_{\rm isol}$ or $\lambda = \lambda_{\rm b}$ solves the distributional Eq. (7) in the main text

$$p_R(r|\lambda) = \sum_{k=0}^{\infty} p_{K^{\rm out}}(k) \int \left( \prod_{j=1}^{k} {\rm d}x_j {\rm d}^2 r_j p_J(x_j) p_R(r_j|\lambda) \right) \delta \left( r - \frac{1}{\lambda} \sum_{j=1}^{k} x_j r_j \right). \tag{S29}$$

The $m+n$-moment of $p_R$ is given by

$$\langle R^n (R^*)^m \rangle = \int {\rm d}^2 r \, p_R(r|\lambda) \, r^n (r^*)^m. \tag{S30}$$

Inserting Eq. (S29) in Eq. (S30) yields

$$\langle R^n (R^*)^m \rangle = \frac{1}{\lambda^n (\lambda^*)^m} \sum_{k=0}^{\infty} p_{K^{\text{out}}}(k) \int \left( \prod_{j=1}^{k} \mathrm{d}x_j \mathrm{d}^2 r_j p_J(x_j) p_R(r_j|\lambda) \right) \left( \sum_{j=1}^{k} x_j r_j \right)^n \left( \sum_{j=1}^{k} x_j r_j^* \right)^m. \tag{S31}$$

Using the multinomial theorem, we can rewrite the above expression as

$$\begin{aligned}\langle R^n (R^*)^m \rangle &= \frac{1}{\lambda^n (\lambda^*)^m} \sum_{k=0}^{\infty} p_{K^{\text{out}}}(k) \sum_{i_1+\cdots+i_k=n} \sum_{j_1+\cdots+j_k=m} C_n(i_1,\ldots,i_k) C_m(j_1,\ldots,j_k) \\ &\quad \times \prod_{t=1}^{k} \langle J_t^{i_t+j_t} \rangle \prod_{t=1}^{k} \left\langle R_t^{i_t} (R_t^*)^{j_t} \right\rangle, \end{aligned} \tag{S32}$$

with the coefficients

$$C_n(i_1,\ldots,i_k) = \frac{n!}{\prod_{t=1}^{k} i_t!}, \quad \text{and} \quad C_m(j_1,\ldots,j_k) = \frac{m!}{\prod_{t=1}^{k} j_t!}. \tag{S33}$$

The sum $\sum_{i_1+\cdots+i_k=n}$ ($\sum_{j_1+\cdots+j_k=m}$) runs over all distinct combinations of non-negative integers $i_1,\ldots,i_k$ ($j_1,\ldots,j_k$) such that $\sum_{t=1}^{k} i_t = n$ ($\sum_{t=1}^{k} j_t = m$). Equation (S32) allows to calculate recursively the moments of $p_R$, as we demonstrate below.

### B. The eigenvalue outlier $\lambda_{\text{isol}}$ and eigenvalues $\lambda_{\text{b}}$ at the boundary of the spectrum

By setting $n = 1$ and $m = 0$ in Eq. (S32), we obtain the fixed-point equation for the mean $\langle R \rangle$

$$\langle R \rangle = \frac{c\langle J \rangle}{\lambda} \langle R \rangle. \tag{S34}$$

Analogously, we can set $n = m = 1$ in Eq. (S32) yielding

$$|\lambda|^2 \langle |R|^2 \rangle = c\langle J^2 \rangle \langle |R|^2 \rangle + \langle K^{\text{out}}(K^{\text{out}} - 1) \rangle \langle J \rangle^2 |\langle R \rangle|^2. \tag{S35}$$

For $n = 2$ and $m = 0$ we get

$$\lambda^2 \langle R^2 \rangle = c\langle J^2 \rangle \langle R^2 \rangle + \langle K^{\text{out}}(K^{\text{out}} - 1) \rangle \langle J \rangle^2 \langle R \rangle^2. \tag{S36}$$

If $\langle R \rangle \neq 0$, then

$$\lambda = \lambda_{\text{isol}} = c\langle J \rangle, \tag{S37}$$

which is the eigenvalue outlier. On the other hand, if $\langle R \rangle = 0$, then

$$\langle |R|^2 \rangle = \frac{c\langle J^2 \rangle}{|\lambda|^2} \langle |R|^2 \rangle, \tag{S38}$$

leading to

$$|\lambda^2| = |\lambda_{\rm b}^2| = c\langle J^2\rangle, \tag{S39}$$

which is the boundary of the spectrum.

### C. The IPR for the right eigenvector associated to an eigenvalue $\lambda_{\rm b}$

The IPR is the ratio between the fourth moment and the second moment of $p_R(r|\lambda)$ (see Eq. (8) in the main text). In this subsection we calculate the IPR for the right eigenvectors associated to $\lambda_{\rm b}$, for which $\langle R\rangle = 0$. We need to distinguish between $\lambda_{\rm b} \in \mathbb{R}$ and $\lambda_{\rm b} \notin \mathbb{R}$ as the distribution $p_R$ is different in each case. This follows from the fact that $\vec{R}(\lambda)$ only contains real components when $\lambda \in \mathbb{R}$.

We first analyze the second moments of $p_R$. If $\lambda_{\rm b} \notin \mathbb{R}$, then Eq. (S36) implies that $\langle R^2\rangle = 0$, while for $\lambda_{\rm b} \in \mathbb{R}$ we have that $\langle R^2\rangle = \langle |R|^2\rangle \neq 0$. We do not need to compute $\langle |R|^2\rangle$, since its value is determined by the choice of normalization for the right eigenvectors.

The third moments $\langle R^3\rangle$ and $\langle R^2 R^*\rangle$ are obtained from Eq. (S32) when $n+m=3$. By inspecting the different terms in Eq. (S32) and taking into account that $\langle R\rangle = \langle R^*\rangle = 0$, we conclude that the third moments solve the linear equations

$$\langle R^3\rangle = \frac{c\langle J^3\rangle}{\lambda_{\rm b}^3}\langle R^3\rangle, \tag{S40}$$

$$\langle R^2 R^*\rangle = \frac{c\langle J^3\rangle}{\lambda_{\rm b}^2 \lambda_{\rm b}^*}\langle R^2 R^*\rangle. \tag{S41}$$

Since $|\lambda_{\rm b}|^2 = c\langle J^2\rangle$, it holds that $\langle R^3\rangle = \langle R^2 R^*\rangle = 0$ for both $\lambda_{\rm b} \in \mathbb{R}$ and $\lambda_{\rm b} \notin \mathbb{R}$.

Lastly, we compute the fourth moments of $p_R$. If $\lambda_{\rm b} \in \mathbb{R}$, then we are only concerned with a single fourth moment since $\langle |R|^4\rangle = \langle R^4\rangle = \langle R^3 R^*\rangle$. In this case, we set $n=4$ and $m=0$ in Eq. (S32), and use the fact that the first and the third moments of $p_R$ are zero, which yields

$$\begin{aligned}\langle R^4\rangle &= \frac{1}{\lambda_{\rm b}^4}\sum_{k=0}^{\infty} p_{K^{\rm out}}(k)\left[k\langle J^4\rangle\langle R^4\rangle + \frac{k(k-1)}{2}\frac{4!}{2!2!}\langle J^2\rangle^2 \langle R^2\rangle^2\right]\\ &= \frac{1}{\lambda_{\rm b}^4}\left[c\langle J^4\rangle\langle R^4\rangle + 3\langle K^{\rm out}(K^{\rm out}-1)\rangle\langle J^2\rangle^2\langle R^2\rangle^2\right].\end{aligned} \tag{S42}$$

Using that $\lambda_{\rm b} = \sqrt{c\langle J^2\rangle}$ in the above expression, we readily obtain from Eq. (S28) the

analytic expression for the IPR

$$\mathcal{I}(\lambda_{\rm b}) = \frac{3\langle J^2\rangle^2 \langle K^{\rm out}(K^{\rm out}-1)\rangle}{c\left(c\langle J^2\rangle^2 - \langle J^4\rangle\right)}, \tag{S43}$$

which is the Eq. (9) in the main text for $\gamma = 2$.

If $\lambda = \lambda_{\rm b} \notin \mathbb{R}$, then we have to consider three different combinations of $n$ and $m$ in Eq. (S32) that yield fourth moments $(n+m=4)$: $\langle R^4\rangle$, $\langle R^3 R^*\rangle$, and $\langle |R|^4\rangle$. Since $\langle R\rangle = \langle R^2\rangle = 0$ for $\lambda_{\rm b} \notin \mathbb{R}$ in the multinomial expansion of Eq. (S32), one obtains that $\langle R^3 R^*\rangle$ and $\langle R^4\rangle$ are determined from the solutions of two linear equations, which have a structure similar to Eqs. (S40) and (S41). From these linear equations for $\langle R^3 R^*\rangle$ and $\langle R^4\rangle$, it is straightforward to verify that $\langle R^3 R^*\rangle = \langle R^4\rangle = 0$ when $|\lambda| = \sqrt{c\langle J^2\rangle}$. We are thus left with a single fourth moment to compute, namely $\langle |R|^4\rangle$. Setting $n = m = 2$ and $\lambda = \lambda_{\rm b} \notin \mathbb{R}$ in Eq. (S32), we get

$$\begin{aligned}\langle |R|^4\rangle &= \frac{1}{|\lambda_{\rm b}|^4} \sum_{k=0}^{\infty} p_{K^{\rm out}}(k) \left[k\langle J^4\rangle\langle |R|^4\rangle + \frac{1}{2}k(k-1)\,2!\,2!\,\langle J^2\rangle^2 \langle |R|^2\rangle^2\right] \\ &= \frac{1}{|\lambda_{\rm b}|^4}\left[c\langle J^4\rangle\langle |R|^4\rangle + 2\langle K^{\rm out}(K^{\rm out}-1)\rangle\langle J^2\rangle^2 \langle |R|^2\rangle^2\right].\end{aligned} \tag{S44}$$

Substituting $|\lambda_{\rm b}| = \sqrt{c\langle J^2\rangle}$ in the above equation, we obtain from Eq. (S28) the analytic expression for the IPR

$$\mathcal{I}(\lambda_{\rm b}) = \frac{\langle |R|^4\rangle}{\langle |R|^2\rangle^2} = \frac{2\langle J^2\rangle^2 \langle K^{\rm out}(K^{\rm out}-1)\rangle}{c\left(c\langle J^2\rangle^2 - \langle J^4\rangle\right)}, \tag{S45}$$

which is the Eq. (9) for $\gamma = 1$.

The only difference between Eqs. (S43) and (S45) is the numerical factor in the numerator, and therefore we can write

$$\frac{\langle |R|^4\rangle}{\langle |R|^2\rangle^2} = \frac{(\gamma+1)\langle J^2\rangle^2 \langle K^{\rm out}(K^{\rm out}-1)\rangle}{c\left(c\langle J^2\rangle^2 - \langle J^4\rangle\right)}, \tag{S46}$$

where $\gamma = 2$ if $\lambda_{\rm b} \in \mathbb{R}$, and $\gamma = 1$ if $\lambda_{\rm b} \notin \mathbb{R}$.

### D. The IPR for the right eigenvector associated to the outlier eigenvalue $\lambda_{\rm isol}$

In this subsection we compute the IPR of the right eigenvector associated to the outlier eigenvalue $\lambda_{\rm isol}$. Since $\langle R\rangle \neq 0$, the the odd moments of $p_R$ are nonzero and the calculation

of the IPR is more involved than for $\lambda = \lambda_{\rm b}$. Substituting $\lambda = \lambda_{\rm isol} = c\langle J \rangle$ in Eq. (S36), we readily obtain the analytic expression [8]

$$\frac{\langle R^2 \rangle}{\langle R \rangle^2} = \frac{\langle J \rangle^2 \langle K^{\rm out}(K^{\rm out} - 1) \rangle}{c \left(c \langle J \rangle^2 - \langle J^2 \rangle\right)}. \tag{S47}$$

Note that the right hand side of the above equation is positive only if $c > c_{\rm gap} = \langle J^2 \rangle / \langle J \rangle^2$, which is the condition for the existence of an outlier (see Eq. (4) in the main text). As before, the value of $\langle R^2 \rangle$ is determined by the choice of the eigenvector normalization.

Combining Eqs. (S47) and (S32), we can calculate the third moment $\langle R^3 \rangle$. Setting $\lambda = \lambda_{\rm isol}$, $n = 3$, and $m = 0$ in Eq. (S32), we find

$$\begin{aligned}
\langle R^3 \rangle &= \frac{1}{\lambda_{\rm isol}^3} \sum_{k=0}^{\infty} p_{K^{\rm out}}(k) \left[ k \langle J^3 \rangle \langle R^3 \rangle + 3k(k-1) \langle J^2 \rangle \langle J \rangle \langle R^2 \rangle \langle R \rangle + \frac{k!}{(k-3)!} \langle J \rangle^3 \langle R \rangle^3 \right] \\
&= \frac{1}{\lambda_{\rm isol}^3} \left[ c \langle J^3 \rangle \langle R^3 \rangle + 3 \langle K^{\rm out}(K^{\rm out} - 1) \rangle \langle J^2 \rangle \langle J \rangle \langle R^2 \rangle \langle R \rangle \right. \\
&\quad \left. + \langle K^{\rm out}(K^{\rm out} - 1)(K^{\rm out} - 2) \rangle \langle J \rangle^3 \langle R \rangle^3 \right],
\end{aligned} \tag{S48}$$

where the combinatorial factor $\frac{k!}{(k-3)!}$ is equal to the number of distinct $k$-dimensional vectors with $k - 3$ components equal to zero and three components equal to one. Using $\lambda_{\rm isol} = c \langle J \rangle$ and dividing the above equation by $\langle R \rangle^3$, leads to

$$\left( c^3 \langle J \rangle^3 - c \langle J^3 \rangle \right) \frac{\langle R^3 \rangle}{\langle R \rangle^3} = 3 \langle K^{\rm out}(K^{\rm out} - 1) \rangle \langle J^2 \rangle \langle J \rangle \frac{\langle R^2 \rangle}{\langle R \rangle^2} + \langle K^{\rm out}(K^{\rm out} - 1)(K^{\rm out} - 2) \rangle \langle J \rangle^3. \tag{S49}$$

Inserting Eq. (S47) in Eq. (S49), we obtain that the third moment is given by

$$\frac{\langle R^3 \rangle}{\langle R \rangle^3} = \frac{3 \langle K^{\rm out}(K^{\rm out} - 1) \rangle^2 \langle J \rangle^3 \langle J^2 \rangle}{\left( c^3 \langle J \rangle^3 - c \langle J^3 \rangle \right) \left( c^2 \langle J \rangle^2 - c \langle J^2 \rangle \right)} + \frac{\langle K^{\rm out}(K^{\rm out} - 1)(K^{\rm out} - 2) \rangle \langle J \rangle^3}{\left( c^3 \langle J \rangle^3 - c \langle J^3 \rangle \right)}. \tag{S50}$$

Lastly, we compute the fourth moment of $p_R$ from the multinomial expansion of Eq. (S32) for $n = 4$ and $m = 0$. The computation is more involved and we need to be careful and take properly into account all combinatorial factors. There are five different types of configurations of non-negative integers $i_1, \ldots, i_k$ that fulfill the constraint $\sum_{t=1}^{k} i_t = 4$. Below we show a single instance of $i_1, \ldots, i_k$ belonging to each type, together with the combinatorial factor arising from taking all possible permutations of a certain configuration

type:

$$i_1 = 4 \quad i_2 = \cdots = i_k = 0 \longrightarrow k,$$
$$i_1 = i_2 = i_3 = i_4 = 1 \quad i_5 = \cdots = i_k = 0 \longrightarrow \frac{k!}{4!(k-4)!},$$
$$i_1 = 3, \ i_2 = 1 \quad i_3 = \cdots = i_k = 0 \longrightarrow k(k-1),$$
$$i_1 = i_2 = 2 \quad i_3 = \cdots = i_k = 0 \longrightarrow \frac{k(k-1)}{2},$$
$$i_1 = 2, \ i_2 = i_3 = 1 \quad i_4 = \cdots = i_k = 0 \longrightarrow \frac{k(k-1)(k-2)}{2}.$$

Taking into account the above combinatorial factors, Eq. (S32) assumes the form

$$\begin{aligned}
\langle R^4 \rangle &= \frac{1}{\lambda_{\text{isol}}^4} \sum_{k=0}^{\infty} p_{K^{\text{out}}}(k) \Big[ k \langle J^4 \rangle \langle R^4 \rangle + k(k-1)(k-2)(k-3) \langle J \rangle^4 \langle R \rangle^4 \\
&\quad + 4k(k-1) \langle J^3 \rangle \langle J \rangle \langle R^3 \rangle \langle R \rangle \\
&\quad + 3k(k-1) \langle J^2 \rangle^2 \langle R^2 \rangle^2 + 6k(k-1)(k-2) \langle J^2 \rangle \langle J \rangle^2 \langle R^2 \rangle \langle R \rangle^2 \Big], \\
&= \frac{1}{\lambda_{\text{isol}}^4} \Big[ c \langle J^4 \rangle \langle R^4 \rangle + \langle K^{\text{out}}(K^{\text{out}}-1)(K^{\text{out}}-2)(K^{\text{out}}-3) \rangle \langle J \rangle^4 \langle R \rangle^4 \\
&\quad + 4 \langle K^{\text{out}}(K^{\text{out}}-1) \rangle \langle J^3 \rangle \langle J \rangle \langle R^3 \rangle \langle R \rangle \\
&\quad + 3 \langle K^{\text{out}}(K^{\text{out}}-1) \rangle \langle J^2 \rangle^2 \langle R^2 \rangle^2 + 6 \langle K^{\text{out}}(K^{\text{out}}-1)(K^{\text{out}}-2) \rangle \langle J^2 \rangle \langle J \rangle^2 \langle R^2 \rangle \langle R \rangle^2 \Big].
\end{aligned}$$

Substituting $\lambda_{\text{isol}} = c \langle J \rangle$ and dividing the above equation by $\langle R \rangle^4$, we arrive at

$$\begin{aligned}
\left( c^4 \langle J \rangle^4 - c \langle J^4 \rangle \right) \frac{\langle R^4 \rangle}{\langle R \rangle^4} &= \langle K^{\text{out}}(K^{\text{out}}-1)(K^{\text{out}}-2)(K^{\text{out}}-3) \rangle \langle J \rangle^4 \\
&\quad + 4 \langle K^{\text{out}}(K^{\text{out}}-1) \rangle \langle J^3 \rangle \langle J \rangle \frac{\langle R^3 \rangle}{\langle R \rangle^3} + 3 \langle K^{\text{out}}(K^{\text{out}}-1) \rangle \langle J^2 \rangle^2 \frac{\langle R^2 \rangle^2}{\langle R \rangle^4} \\
&\quad + 6 \langle K^{\text{out}}(K^{\text{out}}-1)(K^{\text{out}}-2) \rangle \langle J^2 \rangle \langle J \rangle^2 \frac{\langle R^2 \rangle}{\langle R \rangle^2}.
\end{aligned}$$

The analytic expression for $\frac{\langle R^4 \rangle}{\langle R \rangle^4}$ is derived by substituting Eqs. (S47) and (S50) in the above equation leading to

$$\begin{aligned}
\frac{\langle R^4 \rangle}{\langle R \rangle^4} &= \frac{\langle K^{\text{out}}(K^{\text{out}}-1)(K^{\text{out}}-2)(K^{\text{out}}-3) \rangle \langle J \rangle^4}{\left( c^4 \langle J \rangle^4 - c \langle J^4 \rangle \right)} \\
&\quad + \frac{12 \langle K^{\text{out}}(K^{\text{out}}-1) \rangle^3 \langle J \rangle^4 \langle J^3 \rangle \langle J^2 \rangle}{\left( c^4 \langle J \rangle^4 - c \langle J^4 \rangle \right) \left( c^3 \langle J \rangle^3 - c \langle J^3 \rangle \right) \left( c^2 \langle J \rangle^2 - c \langle J^2 \rangle \right)} \\
&\quad + \frac{4 \langle K^{\text{out}}(K^{\text{out}}-1) \rangle \langle K^{\text{out}}(K^{\text{out}}-1)(K^{\text{out}}-2) \rangle \langle J \rangle^4 \langle J^3 \rangle}{\left( c^4 \langle J \rangle^4 - c \langle J^4 \rangle \right) \left( c^3 \langle J \rangle^3 - c \langle J^3 \rangle \right)} \\
&\quad + \frac{3 \langle K^{\text{out}}(K^{\text{out}}-1) \rangle^3 \langle J^2 \rangle^2 \langle J \rangle^4}{\left( c^4 \langle J \rangle^4 - c \langle J^4 \rangle \right) \left( c^2 \langle J \rangle^2 - c \langle J^2 \rangle \right)^2} \\
&\quad + \frac{6 \langle K^{\text{out}}(K^{\text{out}}-1)(K^{\text{out}}-2) \rangle \langle K^{\text{out}}(K^{\text{out}}-1) \rangle \langle J^2 \rangle \langle J \rangle^4}{\left( c^4 \langle J \rangle^4 - c \langle J^4 \rangle \right) \left( c^2 \langle J \rangle^2 - c \langle J^2 \rangle \right)}. \quad (S51)
\end{aligned}$$

The IPR follows from Eq. (S28) (see also Eq. (8) in the main text). Using Eq. (S47), we express $\langle R \rangle^4$ in terms of $\langle R^2 \rangle^2$, obtaining the following analytic expression for the IPR of the right eigenvector associated to $\lambda_{\rm isol}$ (see Eq. (10) in the main text)

$$\frac{\langle R^4 \rangle}{\langle R^2 \rangle^2} = \frac{3\beta_1 \langle J^2 \rangle^2}{(c^4 \langle J \rangle^4 - c \langle J^4 \rangle)} + \frac{\beta_3 \left(c^2 \langle J \rangle^2 - c \langle J^2 \rangle\right)^2}{\beta_1^2 \left(c^4 \langle J \rangle^4 - c \langle J^4 \rangle\right)} + \frac{12\beta_1 \langle J^3 \rangle \langle J^2 \rangle \left(c^2 \langle J \rangle^2 - c \langle J^2 \rangle\right)}{\left(c^4 \langle J \rangle^4 - c \langle J^4 \rangle\right)\left(c^3 \langle J \rangle^3 - c \langle J^3 \rangle\right)}$$
$$+ \frac{4\beta_2 \langle J^3 \rangle \left(c^2 \langle J \rangle^2 - c \langle J^2 \rangle\right)^2}{\beta_1 \left(c^4 \langle J \rangle^4 - c \langle J^4 \rangle\right)\left(c^3 \langle J \rangle^3 - c \langle J^3 \rangle\right)} + \frac{6\beta_2 \langle J^2 \rangle \left(c^2 \langle J \rangle^2 - c \langle J^2 \rangle\right)}{\beta_1 \left(c^4 \langle J \rangle^4 - c \langle J^4 \rangle\right)}, \qquad (S52)$$

where the coefficients $\beta_\ell$ ($\ell = 1, 2, 3$) depend solely on the outdegree distribution and are defined as

$$\beta_\ell = \sum_{k=\ell+1}^{\infty} p_{K^{\rm out}}(k) \frac{k!}{(k-\ell-1)!}. \qquad (S53)$$

## S4. THE HIGH CONNECTIVITY LIMIT FOR THE DISTRIBUTION OF EIGENVECTOR COMPONENTS

In this section, we derive the analytic expressions for the high connectivity limit $c \to \infty$ of $p_R(r|\lambda)$ given by Eqs. (22)-(25) in the main text. We consider explicitly the cases of regular, Poisson, and exponential random graphs, and we discuss why the approach presented here does not work for random graphs with Borel degree distributions.

### A. General formalism

In general, the eigenvector components are complex random variables and $p_R(r|\lambda)$ represents the joint distribution of the real and imaginary components $({\rm Re}(R), {\rm Im}(R))$ in the complex plane. We consider the characteristic function

$$g_R(u, v|\lambda) = \int {\rm d}^2 r \, p_R(r|\lambda) e^{-iu{\rm Re}(r) - iv{\rm Im}(r)} \qquad (S54)$$

of $p_R(r|\lambda)$, where ${\rm d}^2 r \equiv {\rm dRe}(r){\rm dIm}(r)$ and the integral is over the complex plane. If we know the analytic expression for $g_R(u, v|\lambda)$, we obtain $p_R(r|\lambda)$ from

$$p_R(r|\lambda) = \int \frac{{\rm d}u{\rm d}v}{4\pi^2} e^{iu{\rm Re}(r) + iv{\rm Im}(r)} g_R(u, v|\lambda). \qquad (S55)$$

In what follows, we compute $g_R(u, v|\lambda)$ in the limit $c \to \infty$, which is then substituted in the above equation to obtain $p_R(r|\lambda)$.

First, we rewrite $g_R(u,v|\lambda)$ by substituting Eq. (S29) in Eq. (S54), yielding

$$g_R(u,v|\lambda) = \sum_{k=0}^{\infty} p_{K^{\text{out}}}(k) \int \left(\prod_{j=1}^{k} dx_j d^2 r_j p_J(x_j) p_R(r_j|\lambda)\right)$$

$$\times \exp\left[-\mathrm{i}u \sum_{j=1}^{k} x_j \mathrm{Re}\left(\frac{r_j}{\lambda}\right) - \mathrm{i}v \sum_{j=1}^{k} x_j \mathrm{Im}\left(\frac{r_j}{\lambda}\right)\right]$$

$$= \sum_{k=0}^{\infty} p_{K^{\text{out}}}(k) \exp\left[k \ln F(u,v|\lambda)\right]. \tag{S56}$$

If $\lambda \in \mathbb{C}$, then the eigenvector components are complex and the function $F(u,v|\lambda)$ is given by

$$F(u,v|\lambda \in \mathbb{C}) = \int dx \, p_J(x) \int d^2 r \, p_R(r) \exp\left(-\frac{xzr}{2\lambda} + \frac{xz^* r^*}{2\lambda^*}\right), \tag{S57}$$

with $z \equiv u + \mathrm{i}v$, and where $(\ldots)^*$ denotes complex conjugation. On the other hand, if $\lambda \in \mathbb{R}$, then the eigenvector components are distributed on the real line and $F(u,v|\lambda)$ is independent of $v$, i.e.,

$$F(u|\lambda \in \mathbb{R}) = \int dx \, p_J(x) \int d^2 r \, p_R(r) \exp\left(-\frac{\mathrm{i}uxr}{\lambda}\right). \tag{S58}$$

In order to compute the high connectivity limit $c \to \infty$ of $g_R$, we expand $F$ in powers of $1/c$ for $c \gg 1$. By representing the exponential in Eqs. (S57) and (S58) as power-series, we obtain the formal expressions

$$F(u,v|\lambda \in \mathbb{C}) = \sum_{n,m=0}^{\infty} \frac{1}{n!m!} \left(-\frac{z}{2}\right)^n \left(\frac{z^*}{2}\right)^m \frac{\langle J^{n+m}\rangle}{\lambda^n (\lambda^*)^m} \langle R^n (R^*)^m\rangle, \tag{S59}$$

$$F(u|\lambda \in \mathbb{R}) = \sum_{n=0}^{\infty} \frac{(-\mathrm{i}u)^n}{n!} \frac{\langle J^n\rangle}{\lambda^n} \langle R^n\rangle. \tag{S60}$$

Substituting the values $\lambda = \lambda_{\text{isol}} = c\langle J\rangle$ or $|\lambda|^2 = |\lambda_{\mathrm{b}}|^2 = c\langle J^2\rangle$ in Eqs. (S59-S60), we obtain expansions in powers of $1/c$. If the moments $\langle R^n (R^*)^m\rangle$ converge to a finite limit for $c \to \infty$, then we can truncate these expansions at terms of $O(1/c)$, whereas if the moments $\langle R^n (R^*)^m\rangle$ diverge, then we need to consider all terms in the series and know all moments of $p_R$.

From the analytic expressions for the eigenvector moments, presented in the last section, we conclude that the moments of $p_R$ are finite if $\mathrm{var}[K^{\text{out}}] = Bc^\alpha$ ($c \gg 1$) and $\alpha \leq 2$. We explicitly verified that the moments $\langle R^n (R^*)^m\rangle$ ($n+m = 1,2,3,4$) converge to a finite limit if the outdegree distribution $p_{K^{\text{out}}}(k)$ is given by a $\delta$ peak (regular graph), a Poisson, or

an exponential distribution, and also the sixth moment of $p_R(r|\lambda)$ converges to a constant in the high connectivity limit for these examples of degree distributions. Thus, we have compelling evidence that all eigenvector moments have a finite limit for $c \rightarrow \infty$ if $\alpha \leq 2$.

For outdegree distributions with $\alpha > 2$, such as the Borel degree distribution, the analysis of Eqs. (S59) and (S60) in the high connectivity limit is more involved because the eigenvector moments scale with $c$ for $c \gg 1$, and therefore we cannot truncate the series at the first terms. Below we illustrate this more clearly by comparing the first few terms of Eq. (S59) for different degree distributions. From now on, we consider, without loosing generality, that the eigenvectors are normalized as $\langle |R|^2 \rangle = 1$.

### B. High connectivity limit of $p_R$ for $\lambda = \lambda_b \in \mathbb{C}$

Using the results from the previous section and using the notation $F^{(c)}(u, v|\lambda_b) \equiv F(u, v|\lambda_b \in \mathbb{C})$, we can write down the first three nonzero terms of Eq. (S59):

$$F^{(c)}(u, v|\lambda_b) = 1 - \frac{|z|^2}{4c} + \frac{|z|^4 \langle J^4 \rangle}{64 \langle J^2 \rangle^2} \frac{\langle |R|^4 \rangle}{c^2}, \tag{S61}$$

where we substituted $|\lambda|^2 = c\langle J^2 \rangle$.

To proceed further, we need to understand how $\langle |R|^4 \rangle$ behaves for large $c$. From Eq. (S45), it follows that $\lim_{c \rightarrow \infty} \langle |R|^4 \rangle$ attains a finite value if the variance of $p_{K^{\text{out}}}(k)$ scales as $\text{var}[K^{\text{out}}] \propto c^\alpha$ with $\alpha \leq 2$. In particular, substituting the results for the outdegree distributions of the first section in Eq. (S45), we obtain the asymptotic behaviours:

$$\text{Poisson and regular}: \lim_{c \rightarrow \infty} \langle |R|^4 \rangle = 2,$$
$$\text{Exponential}: \lim_{c \rightarrow \infty} \langle |R|^4 \rangle = 4,$$
$$\text{Borel}: \lim_{c \rightarrow \infty} \frac{\langle |R|^4 \rangle}{c} = 2.$$

Hence, for regular, Poisson, and exponential random graphs, for which $\alpha \leq 2$, we can truncate Eq. (S59) at the term of order $O(1/c)$, since the contribution involving $\langle |R|^4 \rangle$ is of the order $O(1/c^2)$. We have also verified that the next term depending on $\langle |R|^6 \rangle$ in Eq. (S59) is of order $O(1/c^3)$ for random graphs with $\alpha \leq 2$. Therefore, for graph ensembles characterized by $\alpha \leq 2$, we set for $c \gg 1$

$$F^{(c)}(u, v|\lambda_b) = 1 - \frac{|z|^2}{4c}. \tag{S62}$$

416In the case of the Borel degree distribution, for which $\alpha > 2$, a more refined analysis of Eq. (S59) is needed since the term depending on $\langle |R|^4 \rangle$ in Eq. (S61) is also of the order $O(1/c)$. Therefore, we cannot truncate the series at the second term. We leave the analysis of the high connectivity limit of $p_R$ for graph ensembles with $\alpha > 2$ for a future work.

Substituting Eq. (S62) in Eq. (S56) and expanding the logarithm for $c \gg 1$, we obtain an expression for $g_R^{(c)}(u,v|\lambda_b) \equiv g_R(u,v|\lambda_b \in \mathbb{C})$

$$g_R^{(c)}(u,v|\lambda_b) = \sum_{k=0}^{\infty} p_{K^{\text{out}}}(k) \exp\left[-\frac{k|z|^2}{4c}\right]. \tag{S63}$$

The above equation depends only on the degree distribution and is independent of $p_J$. Consequently, at this point we have to specify $p_{K^{\text{out}}}(k)$ in order to explicitly perform the summation over the degrees. For regular, Poisson and exponential outdegree distributions (see Eqs. (S2), (S5) and (S10)), we explicitly sum the series and take the limit $c \to \infty$, obtaining

$$g_R^{(c)}(u,v|\lambda_b) = \exp\left(-\frac{|z|^2}{4}\right) \tag{S64}$$

for Poissonian and regular degree distributions, and

$$g_R^{(c)}(u,v|\lambda_b) = \left(1 + \frac{|z|^2}{4}\right)^{-1} \tag{S65}$$

for exponential degree distributions.

Lastly, we insert Eq. (S64) in Eq. (S55) and calculate the Gaussian integral over $u$ and $v$ to obtain the expression

$$p_R^{(c)}(r|\lambda_b) = \frac{1}{\pi} e^{-|r|^2}, \tag{S66}$$

valid for degree distributions with $\alpha < 2$. The above equations is the Eq. (22) in the main text. The above result is consistent with the standard prediction of random matrix theory according to which the eigenvector components of fully-connected random graphs with Gaussian distributed edges are Gaussian distributed random variables [14]. From Eq. (S66), it is straightforward to derive the so-called Porter-Thomas distribution for the eigenvector amplitudes $\{|R_i|^2\}_{i=1,\ldots,N}$ [14]. Although it is natural to expect that the spectral properties of random graphs converge to those of Gaussian random matrices in the limit $c \to \infty$, Eq. (S63) makes clear that the high connectivity limit of $p_R(r|\lambda_b)$ depends on the degree distribution.

As an example of graph ensemble for which the high connectivity limit of $p_R$ does not converge to the Porter-Thomas distribution of random matrix theory, we compute $p_R$ for an exponential degree distribution ($\alpha = 2$). Substituting Eq.(S65) in Eq. (S55), we obtain

$$p_R^{(c)}(r|\lambda_\text{b}) = 4\int_{-\infty}^{\infty} \frac{\text{d}u\text{d}v}{4\pi^2} e^{iu\text{Re}(r)+iv\text{Im}(r)} \frac{1}{4 + u^2 + v^2}. \tag{S67}$$

After changing the integration variables in Eq. (S67) to polar coordinates $(\rho, \theta)$, i.e., $u = \rho\cos\theta$ and $v = \rho\sin\theta$, we can integrate over $\theta$ and find

$$p_R^{(c)}(r|\lambda_\text{b}) = \frac{2}{\pi} \int_0^\infty \text{d}\rho \frac{\rho}{(4+\rho^2)} J_0\left(\rho|r|\right), \tag{S68}$$

where $J_0(x)$ is a Bessel function of the first kind. The integral over $\rho$ is calculated through an integration by parts, leading to the final result

$$p_R^{(c)}(r|\lambda_\text{b}) = \frac{2}{\pi} K_0\left(2|r|\right), \tag{S69}$$

where $K_0^{\text{out}}(x)$ is a modified Bessel function of the second kind [15]. The above equation is Eq. (24) in the main text.

### C.  High connectivity limit of $p_R$ for $\lambda = \lambda_\text{b} \in \mathbb{R}$

Substituting $\lambda = \pm\sqrt{c\langle J^2\rangle}$ in Eq. (S60), we obtain the following expression for $F^{(r)}(u|\lambda_\text{b}) \equiv F(u|\lambda_\text{b} \in \mathbb{R})$ when $c \gg 1$

$$F^{(r)}(u|\lambda_\text{b}) = 1 - \frac{u^2}{2c}, \tag{S70}$$

provided $p_{K^{\text{out}}}(k)$ is such that $\alpha \leq 2$. Inserting Eq. (S70) in Eq. (S56) and expanding the logarithm for large $c$, we find

$$g_R^{(r)}(u|\lambda_\text{b}) = \sum_{k=0}^{\infty} p_{K^{\text{out}}}(k) \exp\left[-\frac{ku^2}{2c}\right], \tag{S71}$$

where we introduced the notation $g_R^{(r)}(u|\lambda_\text{b}) \equiv g_R(u|\lambda_\text{b} \in \mathbb{R})$. If $p_{K^{\text{out}}}(k)$ is a Poisson or regular distribution (or more generally $\alpha < 2$), then

$$g_R^{(\text{r})}(u|\lambda_\text{b}) = \exp\left(-\frac{u^2}{2}\right), \tag{S72}$$

while for the exponential degree distribution with $\alpha = 2$ we obtain

$$g_R^{(\text{r})}(u|\lambda_\text{b}) = \left(1 + \frac{u^2}{2}\right)^{-1}. \tag{S73}$$

Lastly, we calculate the integral in Eq. (S55). Substituting Eq. (S72) in Eq. (S55) and calculating the Gaussian integrals over $u$ and $v$, we obtain the following result for $p_R^{(r)}(r|\lambda_{\rm b}) = p_R^{(r)}(r|\lambda_{\rm b} \in \mathbb{R})$

$$p_R^{(r)}(r|\lambda_{\rm b}) = \delta\left[{\rm Im}(r)\right] \frac{1}{\sqrt{2\pi}} e^{-\frac{1}{2}[{\rm Re}(r)]^2}, \tag{S74}$$

valid for $\alpha < 2$. Now we turn our attention to the exponential degree distribution for which $\alpha = 2$. Substituting Eq. (S73) in Eq. (S55), we get

$$p_R^{(r)}(r|\lambda_{\rm b}) = 2\delta\left[{\rm Im}(r)\right] \int_{-\infty}^{\infty} \frac{{\rm d}u}{2\pi} e^{iu{\rm Re}(r)} \frac{1}{2+u^2}, \tag{S75}$$

which leads to the exponential distribution

$$p_R^{(r)}(r|\lambda_{\rm b}) = \frac{1}{\sqrt{2}} \delta\left[{\rm Im}(r)\right] e^{-\sqrt{2}|{\rm Re}(r)|}. \tag{S76}$$

### D. High connectivity limit of $p_R$ for $\lambda = \lambda_{\rm isol}$

We set $\lambda = \lambda_{\rm isol} = c\langle J\rangle$ in Eq. (S60) and truncate the series at the second term, yielding

$$F(u|\lambda_{\rm isol}) = 1 - \frac{iu\langle R\rangle}{c}. \tag{S77}$$

Analogous to the case $\lambda = \lambda_{\rm b}$, the moments $\langle R^n \rangle$ $(n = 1, 2, 3, \dots)$ of the eigenvector distribution attain a finite value in the limit $c \to \infty$ if $\alpha \le 2$ and $\lambda = \lambda_{\rm isol}$. This ensures that we can truncate the series in Eq. (S60) at the second term. Substituting Eq. (S47) in Eq. (S77) and setting the normalization $\langle R^2 \rangle = 1$, we obtain

$$F(u|\lambda_{\rm isol}) = 1 - \frac{iu}{\sqrt{\langle K^{\rm out}(K^{\rm out} - 1)\rangle}}, \tag{S78}$$

which leads to the following expression for the characteristic function

$$g_R(u|\lambda_{\rm isol}) = \sum_{k=0}^{\infty} p_{K^{\rm out}}(k) \exp\left[-\frac{iuk}{\sqrt{\langle K^{\rm out}(K^{\rm out} - 1)\rangle}}\right]. \tag{S79}$$

By substituting the variance ${\rm var}[K^{\rm out}] = Bc^\alpha$ in the above equation, we obtain the Eq. (21) appearing in the main text. For $\alpha < 2$, which includes regular and Poisson degree distributions, Eq. (S79) yields for $c \to \infty$

$$g_R(u|\lambda_{\rm isol}) = e^{-iu}, \tag{S80}$$

while, for the exponential degree distribution, Eq. (S79) yields for $c \to \infty$

$$g_R(u|\lambda_{\text{isol}}) = \frac{\sqrt{2}}{\sqrt{2} + iu}. \tag{S81}$$

Finally, in order to obtain $p_R(r|\lambda_{\text{isol}})$, we substitute Eqs. (S80) and (S81) in Eq. (S55), solve the remainder integrals, and take the limit $c \to \infty$, obtaining the following result for $\alpha < 2$

$$p_R(r|\lambda_{\text{isol}}) = \delta\left[\text{Im}(r)\right] \delta\left[\text{Re}(r) - 1\right]. \tag{S82}$$

For the exponential degree distribution with $\alpha = 2$, we get

$$p_R(r|\lambda_{\text{isol}}) = \sqrt{2}\,\delta\left[\text{Im}(r)\right] \Theta\left[\text{Re}(r)\right] e^{-\sqrt{2}\text{Re}(r)}, \tag{S83}$$

where $\Theta(\ldots)$ denotes the Heaviside step function. Eqs. (S82) and (S83) are, respectively, the Eqs. (23) and (25) in the main text, which we aimed to derive. Equations (S69), (S76), (S82), and (S83) explicitly show that the high connectivity limit of the eigenvector distributions at $\lambda = \lambda_{\text{b}}$ and $\lambda = \lambda_{\text{out}}$ is generally not given by the predictions of random matrix theory [14].

## S5. COMPARISON WITH DIRECT DIAGONALIZATION RESULTS

We compare the theoretical results for the IPR of $\vec{R}(\lambda_{\text{b}})$ and $\vec{R}(\lambda_{\text{isol}})$, given by Eqs. (9) and (10) in the main text, with results for the IPR obtained from direct diagonalization of adjacency matrices with finite size $n$.

In order to compare the theory with direct diagonalization, we order the eigenvalues of random matrices $\mathbf{A}$ of size $n \times n$ in the following way:

$$|\lambda_1| \geq |\lambda_2| \geq \ldots \geq |\lambda_n|. \tag{S84}$$

If $|\lambda_j| = |\lambda_{j+1}|$, then we order the eigenvalues such that $\text{Im}(\lambda_j) > \text{Im}(\lambda_{j+1})$. Given the above ordering, we define the relative rank of an eigenvalue $\lambda_j$ as

$$y \equiv \frac{j}{n}. \tag{S85}$$

The eigenvalue outlier $\lambda_{\text{isol}}$ and an eigenvalue $\lambda_{\text{b}}$ located at the boundary of the spectrum both have a relative rank that satisfies $y \to 0$.

The inverse participation ratio associated with the right eigenvector of $\lambda_j$ is given by

$$\mathcal{I}(\lambda_j) = \frac{n \sum_{i=1}^{n} |R_i(\lambda_j)|^4}{\left(\sum_{i=1}^{n} |R_i(\lambda_j)|^2\right)^2}. \tag{S86}$$

In the limit $n \to \infty$, we consider the function

$$\mathcal{I}_y = \lim_{n \to \infty} \mathcal{I}(\lambda_{ny}) \tag{S87}$$

defined on the interval $[0, 1]$, which is a deterministic variable if $\mathcal{I}(\lambda_j)$ is self-averaging. For finite $n$, we are interested in the mean value

$$\langle \mathcal{I}(\lambda_j) \rangle = \left\langle \frac{n \sum_{i=1}^{n} |R_i(\lambda_j)|^4}{\left(\sum_{i=1}^{n} |R_i(\lambda_j)|^2\right)^2} \right\rangle \tag{S88}$$

and in the distribution $p_{\mathcal{I}(\lambda_j)}$ for the IPR

$$p_{\mathcal{I}(\lambda_j)}(x) = \left\langle \delta(x - \mathcal{I}(\lambda_j)) \right\rangle. \tag{S89}$$

In the first subsection, we present direct diagonalization results for the mean IPR of right eigenvectors, given by Eq. (S88), as a function of the relative rank $y$, and we show that the theoretical results derived in the main text apply to right eigenvectors with rank $y \approx 0$. In the second subsection, we compute the distribution of the IPR given by Eq. (S89). Finally, in the last subsection, we plot the mean IPR as a function of $c$ for right eigenvectors with small rank $j = 1, 2, 3, 4, 5$ and we compare these results with the Fig. 2 in the main text.

### A. IPR as a function of the rank

The properties of $\mathcal{I}(\lambda_j)$ depend whether $\lambda_j \notin \mathbb{R}$ or $\lambda_j \in \mathbb{R}$, since in the former case the right eigenvectors have complex entries, while in the latter case the eigenvectors have real entries. Therefore, it is convenient to consider the conditional averages

$$\langle \mathcal{I}(\lambda_j) | \lambda_j \notin \mathbb{R} \rangle = \left\langle \frac{n \sum_{i=1}^{n} |R_i(\lambda_j)|^4}{\left(\sum_{i=1}^{n} |R_i(\lambda_j)|^2\right)^2} \bigg| \lambda_j \notin \mathbb{R} \right\rangle, \tag{S90}$$

and

$$\langle \mathcal{I}(\lambda_j) | \lambda_j \in \mathbb{R} \rangle = \left\langle \frac{n \sum_{i=1}^{n} |R_i(\lambda_j)|^4}{\left(\sum_{i=1}^{n} |R_i(\lambda_j)|^2\right)^2} \bigg| \lambda_j \in \mathbb{R} \right\rangle. \tag{S91}$$

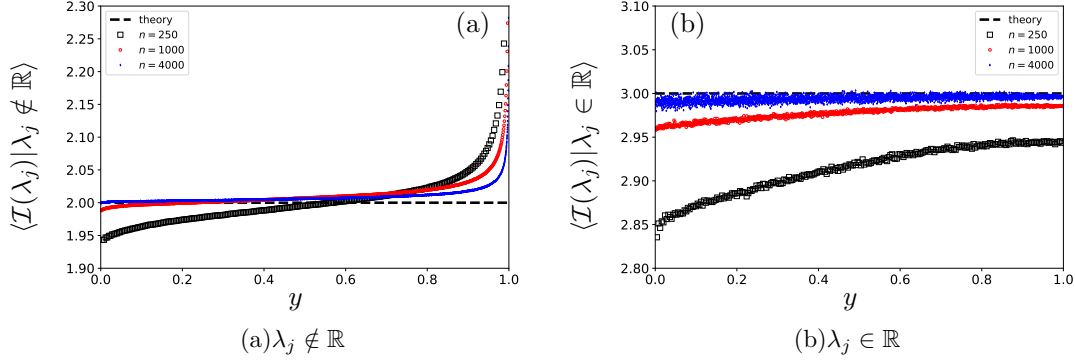

(a) $\lambda_j \notin \mathbb{R}$

(b) $\lambda_j \in \mathbb{R}$

FIG. S2: The average inverse participation ratios $\langle \mathcal{I}(\lambda_j)|\lambda_j \notin \mathbb{R}\rangle$ [Panel (a)] and $\langle \mathcal{I}(\lambda_j)|\lambda_j \in \mathbb{R}\rangle$ [Panel (b)] as a function of $y = j/n$ for a random, regular, and directed graph with $c = 5$ and $p_J(x) = \delta(x-1)$. Empirical estimates (markers) are compared with the theoretical results (dashed lines) valid for $y \to 0$ and given by Eq. (S98). Markers for $n = 250$ (black squares) and $n = 1000$ (red circles) are sample means over $1e+6$ matrix realizations, and markers for $n = 4000$ (blue diamonds) are sample means over $1e+5$ matrix realizations.

Thus, in the limit $n \to \infty$, we study the functions

$$\mathcal{I}_y^{(c)} = \lim_{n\to\infty} \langle \mathcal{I}(\lambda_{ny})|\lambda_{ny} \notin \mathbb{R}\rangle \tag{S92}$$

and

$$\mathcal{I}_y^{(r)} = \lim_{n\to\infty} \langle \mathcal{I}(\lambda_{ny})|\lambda_{ny} \in \mathbb{R}\rangle, \tag{S93}$$

which are defined on the interval $[0,1]$.

According to the theory presented in the main text, we have that (see Eq. (9))

$$\lim_{y\to 0} \mathcal{I}_y^{(c)} = \frac{2\left[\langle (K^{\text{out}})^2\rangle - c\right]}{c\left(c - \langle J^4\rangle/\langle J^2\rangle^2\right)} \tag{S94}$$

and

$$\lim_{y\to 0} \mathcal{I}_y^{(r)} = \frac{3\left[\langle (K^{\text{out}})^2\rangle - c\right]}{c\left(c - \langle J^4\rangle/\langle J^2\rangle^2\right)}, \tag{S95}$$

provided $\mathcal{I}_y^{(c)}$ and $\mathcal{I}_y^{(r)}$ are self-averaging, and given that $\lambda = \lambda_b$. The quantities $\mathcal{I}_y^{(c)}$ and $\mathcal{I}_y^{(r)}$ are self-averaging if $p_J(x) = \delta(x-1)$, as discussed in Ref. [13]. If $p_J(x)$ has a finite variance, then Eqs. (S94- S94) apply to a very good approximation, and they are exact if the average values are replaced by the typical values in the right-hand side of Eqs. (S92-S93) [13].

In the special case of the adjacency matrices of directed random graphs such that

$$p_{K^{\text{out}}}(k) = \delta_{k,c} \tag{S96}$$

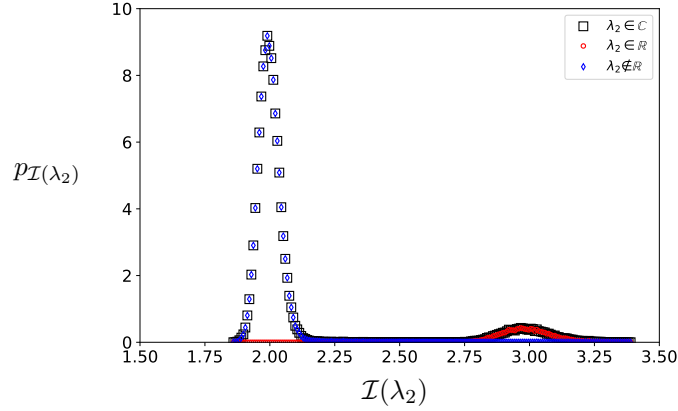

FIG. S3: Distribution $p_{\mathcal{I}(\lambda_2)}$ of $\mathcal{I}(\lambda_2)$ for random, directed, regular graphs with $c = 5$ and $p_J(x) = \delta(x-1)$ (black squares). The distribution is constructed from $1e+5$ matrix realizations [same data as in Fig. S2]. We also plot the distribution of $\mathcal{I}(\lambda_2)$ when $\lambda_2$ is conditioned to be non-real (red circles) and real (blue diamonds).

and

$$p_J(x) = \delta(J-1), \tag{S97}$$

we obtain

$$\lim_{y \to 0} \mathcal{I}^{(c)}(y) = 2, \quad \lim_{y \to 0} \mathcal{I}^{(r)}(y) = 3, \tag{S98}$$

independent of the average degree $c$.

In Fig. S2, we estimate $\langle \mathcal{I}(\lambda_{ny}) | \lambda_{ny} \notin \mathbb{R} \rangle$ and $\langle \mathcal{I}(\lambda_{ny}) | \lambda_{ny} \in \mathbb{R} \rangle$ as a function of the relative rank $y$ using sample means obtained from direct diagonalization results of directed random graphs with $p_{K^{\text{out}}}(k) = \delta_{k,c}$, $p_J(x) = \delta(x-1)$, and $c = 5$, and for three values of the system size: $n = 250, 1000, 4000$. Figure S2 compares the numerical estimates for the mean IPR's at $y \approx 0$ with the theoretical expressions given by Eq. (S98). We observe that Eq. (S98) is well corroborated by the numerical diagonalization results.

In addition, we make a couple of interesting empirical observations. From Fig. 2(b) it is clear that

$$\mathcal{I}^{(r)}(y) = 3 \tag{S99}$$

for any value of $y \in [0, 1]$. On the other hand, from Fig. 2(a) it is not clear what is the value of $\mathcal{I}^{(c)}(y)$ for $y > 0$ [note that the theory for the IPR Eq. (9) only applies at the boundary of the spectrum, i.e. for $y = 0$].

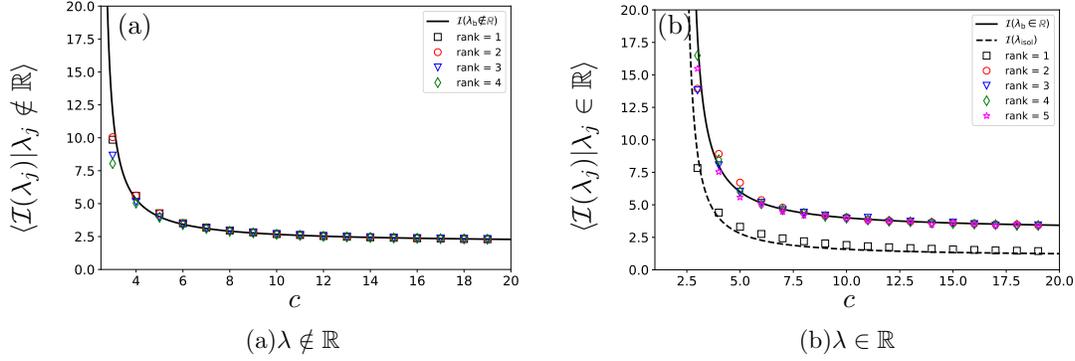

(a) $\lambda \notin \mathbb{R}$

(b) $\lambda \in \mathbb{R}$

FIG. S4: Sample means for the IPR of right eigenvectors in the adjacency matrices of random directed graphs with $p_{K^{\rm out}}$ a Poisson distribution of mean $c$, $p_J$ a Gaussian distribution with mean $\mu = 1$ and variance $\sigma^2 = 1$, and $n = 1000$. Panel (a): each marker is the sample mean over 1e+3 matrix realizations. The solid line denotes the theoretical value $\mathcal{I}(\lambda_{\rm b})$ in Eq. (S103) with $\gamma = 2$. Panel (b): each marker is the sample mean over 1e+4 matrix realizations. The solid line denotes the theoretical value $\mathcal{I}(\lambda_{\rm b})$ in Eq. (S103) with $\gamma = 3$ and the dashed line is the theoretical value $\mathcal{I}(\lambda_{\rm isol})$ in Eq. (S104). Note that for $c > c_{\rm gap} = 2$ (see Eq.(4)) the eigenvalue of rank 1 will be the real-valued outlier, and therefore in Panel (a) we have not considered rank 1 eigenvalues.

### B.  Distribution of the IPR and self-averaging

In Fig. S3, we plot numerical estimates for the distribution of $\mathcal{I}(\lambda_2)$, given by Eq. (S89), for the eigenvalue of rank $j = 2$ of adjacency matrices of directed random graphs with $p_{k^{\rm out}}(k) = \delta_{k,c}$, $p_J(x) = \delta(x-1)$, and $c = 5$. We observe that the distribution is peaked around two values, namely, 2 and 3. These two values correspond with $\mathcal{I}(\lambda_2)$ at $\lambda_2 \notin \mathbb{R}$ and $\lambda_2 \in \mathbb{R}$, respectively. This is evident from plotting the distributions conditioned on either $\lambda_2 \notin \mathbb{R}$ or $\lambda_2 \in \mathbb{R}$. These results are consistent with the theoretical values of $\lambda_{\rm b} \notin \mathbb{R}$ and $\lambda_{\rm b} \in \mathbb{R}$ in Eq. (S98), and hence corroborate the theory for the IPR of directed random graphs presented in the main paper.

In the limit of large $n$, the height of the peak centred at the value $\mathcal{I}(\lambda_2) = 3$ will converge to zero for $n \to \infty$. This is because the number of real eigenvalues of the adjacency matrix of a regular random graph scales as $\sqrt{n}$, as shown in Figure 3 of Ref. [9]. Moreover, in the limit of large $n$, the widths of the two peaks in Fig. S3 will converge to zero.

## C. Localization transition

Lastly, we show numerical evidence from direct diagonalization results for the localization transition of right eigenvectors in the adjacency matrix of directed random graphs.

In Fig. S4 we present empirical estimates for $\langle \mathcal{I}(\lambda_j) | \lambda_j \notin \mathbb{R} \rangle$ and $\langle \mathcal{I}(\lambda_j) | \lambda_j \in \mathbb{R} \rangle$ for eigenvalues of rank $j = 1, 2, 3, 4, 5$ in random directed graphs with

$$p_{K^{\rm out}}(k) = \frac{c^k e^{-c}}{k!} \tag{S100}$$

and with

$$p_J(x) = \frac{1}{\sqrt{2\pi\sigma^2}} e^{-\frac{(x-\mu)^2}{2\sigma^2}} \tag{S101}$$

with $\mu = 1$ and $\sigma^2 = 1$. Note that in the limit $n \to \infty$ graphs with a Poisson outdegree distribution are equivalent to directed Erdős-Rényi graphs for which the $C_{ij}$ are i.i.d. random variables drawn from the distribution

$$p_{C_{ij}}(x) = \frac{c}{n}\delta_{x,1} + \left(1 - \frac{c}{n}\right)\delta_{x,0}. \tag{S102}$$

Fig. S4 should be compared with Fig. 2 in the main text that plots the theoretical expressions in the limit $n \to \infty$ given by Eqs. (9) and (10). In the present case,

$$\mathcal{I}(\lambda_{\rm b}) = \frac{(\gamma + 1)c}{c - 5/2} \tag{S103}$$

and

$$\mathcal{I}(\lambda_{\rm isol}) = \frac{c}{c^3 - 10}\left\{3 + (c-1)\left[5 + c + \frac{16}{c-2}\right]\right\}. \tag{S104}$$

We observe in Fig. S4 that the theoretical expressions given by Eqs. (S103) and (S104) are very well corroborated by direct diagonalization results. This shows that the theory for the IPR presented in the main text works well even for matrices of relative small size $n = 1000$.

[1] M. E. J. Newman, S. H. Strogatz, and D. J. Watts, Phys. Rev. E **64**, 026118 (2001), URL https://link.aps.org/doi/10.1103/PhysRevE.64.026118.

[2] B. Bollobás and B. Béla, *Random graphs*, 73 (Cambridge university press, 2001).

[3] A. Dembo, A. Montanari, et al., Brazilian Journal of Probability and Statistics **24**, 137 (2010).

[4] M. Newman, *Networks: An Introduction* (OUP Oxford, 2010), ISBN 9780199206650, URL https://books.google.com.br/books?id=q7HVtpYVfC0C.

[5] S. N. Dorogovtsev and J. F. Mendes, *Evolution of networks: From biological nets to the Internet and WWW* (OUP Oxford, 2013).

[6] B. K. Fosdick, D. B. Larremore, J. Nishimura, and J. Ugander, SIAM Review **60**, 315 (2018), https://doi.org/10.1137/16M1087175, URL https://doi.org/10.1137/16M1087175.

[7] B. Bollobs, European Journal of Combinatorics **1**, 311 (1980), ISSN 0195-6698, URL http://www.sciencedirect.com/science/article/pii/S0195669880800308.

[8] I. Neri and F. L. Metz, Phys. Rev. Lett. **117**, 224101 (2016).

[9] F. L. Metz, I. Neri, and T. Rogers, Journal of Physics A: Mathematical and Theoretical **52**, 434003 (2019).

[10] P. van Mieghem, *Graph Spectra for Complex Networks* (Cambridge University Press, 2012), ISBN 9781107411470, URL https://books.google.com.br/books?id=hwJFMAEACAAJ.

[11] E. Borel, C. R. Acad. Sci. **214**, 452 (1942).

[12] J. C. Tanner, Biometrika **48**, 222 (1961), ISSN 0006-3444, https://academic.oup.com/biomet/article-pdf/48/1-2/222/607732/48-1-2-222.pdf, URL https://doi.org/10.1093/biomet/48.1-2.222.

[13] I. Neri and F. L. Metz, Phys. Rev. Research **2**, 033313 (2020), URL https://link.aps.org/doi/10.1103/PhysRevResearch.2.033313.

[14] A. D. Mirlin, Physics Reports **326**, 259 (2000), ISSN 0370-1573, URL http://www.sciencedirect.com/science/article/pii/S0370157399000915.

[15] I. Gradshteyn and I. Ryzhik, *Table of Integrals, Series, and Products* (Elsevier Science, 2014), ISBN 9781483265643, URL https://books.google.com.br/books?id=F7jiBQAAQBAJ.



 \end{document}